\documentclass[usenatbib,twocolumn]{mn2e}
\usepackage{graphicx,verbatim,amsmath,amssymb,enumitem}
\begin{document}
\topmargin-1cm

\title{Characterizing simulated galaxy stellar mass histories}

\author[Cohn and van de Voort]{J. D. Cohn${}^{1,2}$\thanks{E-mail: jcohn@berkeley.edu}, Freeke van de
  Voort${}^{2,3}$\\
${}^1$Space Sciences Laboratory 
  University of California, Berkeley, CA 94720, USA\\
${}^2$Theoretical Astrophysics Center,
  University of California, Berkeley, CA 94720, USA\\
${}^3$Academia Sinica Institute of Astronomy and Astrophysics,
P.O. Box 23-141, Taipei 10617, Taiwan}

\date{Received \today; in original form 11 June 2014}

\maketitle
\label{firstpage}

\begin{abstract}
  Cosmological galaxy formation simulations can now produce rich and
  diverse ensembles of galaxy histories.  These simulated galaxy
  histories, taken all together, provide an answer to the question
  ``how do galaxies form?'' for the models used to construct them.  We
  characterize such galaxy history ensembles both to understand their
  properties and to identify points of comparison for histories within
  a given galaxy formation model or between different galaxy formation
  models and simulations.  We focus primarily on stellar mass
  histories of galaxies with the same final stellar mass, for six
  final stellar mass values and for three different simulated galaxy
  formation models (a semi-analytic model built upon the dark matter
  Millennium simulation and two models from the hydrodynamical
  OverWhelmingly Large Simulations project).  Using principal
  component analysis (PCA) to classify scatter around the average
  stellar mass history, we find that one fluctuation dominates for all
  sets of histories we consider, although its shape and contribution
  can vary between samples.  We correlate the PCA characterization
  with several $z=0$ galaxy properties (to connect with survey
  observables) and also compare it to some other galaxy history
  properties.  We then explore separating galaxy stellar mass
  histories into classes, using the largest PCA contribution, k-means
  clustering, and simple Gaussian mixture models.  For three component
  models, these different methods often gave similar results.  These
  history classification methods provide a succinct and often quick
  way to characterize changes in the full ensemble of histories of a
  simulated population as physical assumptions are varied, to compare
  histories of different simulated populations to each other, and to
  assess the relation of simulated histories to fixed time
  observations.
\end{abstract}


\section{Introduction}
In spite of the complexity of galaxy evolution and the diversity of
individual galaxies, patterns have emerged in galaxy formation.  For 
example, roughly speaking, galaxies are expected to start out star 
forming and small and then grow larger to eventually become ``red and 
dead''.  This qualitative pattern provides targets for theories to 
predict and explain, both qualitatively and, with more detailed theories and
observations, quantitatively. 
 In this paper we look for patterns in
populations of simulated
galaxy stellar mass histories, using a few standard statistical
methods.  The aim is to find characterizations of
sets of galaxy histories produced by simulations, which can be used to
intercompare different sets of full histories, or different simulation
models.
We focus on simulations, because simulations produce histories directly.
Observationally one can estimate how populations at different times
are connected, but in a
simulation one can watch directly how populations evolve into each other.
The full set of histories from a simulation explores the 
consequences of
the assumptions (e.g. relations and sub-grid models) 
used in their construction.  In fact,
the full set of histories produced by a simulation provide one
answer to the question
``how do galaxies form?'' (for a particular model).

Given a set of simulated or observationally inferred galaxy histories,
an immediate quantity of interest is the
average history.
Average histories of  some galaxy properties have been deduced using
a combination of models and observations
(for average stellar mass histories, e.g., \citet{ConWec09,BehWecCon13a,BehWecCon13b},  
for the star forming main sequence, e.g., \citet{Oli10,Kar11,Lei12}
and for star
formation histories, e.g. \citet{MadDic14,MitLacColBau14,Sim14}).
As observational and simulation data grow in amount and detail,
samples and questions can be refined beyond such average properties.
One extension is to track subpopulations,
either directly in simulations or observationally via the use of
various proxies (e.g. the most massive galaxies). 
 
Here we take a different route,
taking all simulated galaxies which share essentially the same final
stellar mass, and classifying properties of the combined set of their
full histories.
These classifications can search for and succinctly capture simplifying 
features of the full simulated galaxy history population.   As such, they
can
be used to identify
points of contact between observations and theoretical
assumptions, to compare theoretical models to each other, and to
compare subsamples within the same theoretical model. 

Our main focus is on ensembles of stellar mass histories.
These are
a priori fairly complicated, as many different physical quantities
and processes can impact the stellar mass.
For instance, stellar mass can grow through star formation
or by merging, and can decrease by, e.g., tidal stripping, stellar
winds or supernovae.  These effects can interact with each other and
with other galaxy properties, complicating the interpretation of 
the consequences of modeling assumptions.
By having a simple description of the full population of stellar mass histories,
consequences of the full interplay between different contributing factors may be easier to identify.

Our primary approach is principal component analysis (PCA), a standard
classification algorithm, although we also briefly explore mixture
model classifications and k-means classifications.  We consider both semi-analytic
models grafted onto a dark matter simulation (Millennium,
\citet{Spr05,Lem06,Guo13}, see \S\ref{sec:methods} below) and two
hydrodynamic simulations from the OverWhelmingly Large Simulations
(OWLS, \citealt{OWLS}) project.  To anticipate some key points, PCA
characterizes fluctuations around an average history and we find that
the histories do in fact, for all cases we consider, roughly peak
around an average history (although not necessarily in a Gaussian
distribution).  Interestingly, we also find that for every sample, the
leading PCA fluctuation significantly dominates the variance around
the average stellar mass history, with close to 90\% of the variance
captured by the first three PCA fluctuations.  This is a large
simplification of the diverse population of galaxy stellar mass
histories.

The total contribution of each fluctuation (i.e.\ principal component) 
around the average history, the distribution of their coefficients and
their shapes carry information about the full ensemble of galaxy stellar
mass histories.
The leading principal component is similar
for several of the Millennium histories and one
corresponding OWLS model.
The amount of variance due to
the leading principal component changes with final stellar mass,
as does the contribution of the leading principal component to
the star forming main sequence inferred from observational samples.
With an eye to determining stellar mass history properties from fixed time
observations, 
we relate some of our classification properties to a range of $z=0$
galaxy observables.  We also find correlations between PCA quantities
and a variety of galaxy halo history and stellar mass history properties used
in other contexts.

The dominance of the first few $PC_n$ allows a simplified
description of the set of all histories sharing the same final stellar
mass, based upon properties of the ensemble of their histories.
We go beyond this basic PCA classification in a few ways.
We briefly explore two other classification methods,
k-means clustering and mixture models, where instead of 
classifying fluctuations
around a single average history, one assigns galaxies to
(usually one of) several characteristic
histories, removing the assumption of symmetry around an average
path. 
We also consider how well the PCA description provides
approximations to stellar mass histories of individual galaxies, when using only the leading principal components.

Finally, we also consider PCA of a few other galaxy history properties.
Closely related to this work, PCA of dark matter halo mass histories 
was performed by \citet{WonTay11}, who found relations to structural
properties of halos such as concentration. We touch upon some
relations between halo history PCA  and stellar mass history PCA
for our stellar mass selected samples.

In \S \ref{sec:methods} the simulations and our application of
principal component analysis, PCA,
are described.  PCA is applied to the
stellar mass histories in \S \ref{sec:pca} 
(applications to star formation rate histories, specific star
formation rate histories and halo histories are briefly mentioned).
In \S \ref{sec:corrlns} correlations of PCA properties with some
galaxy $z=0$ and galaxy and halo history properties are shown, while in \S
\ref{sec:splitsamp} we describe our different separations of the full
sample into classes by using k-means clustering and mixture models.
We discuss and summarize our results in \S \ref{sec:discussion}. The
Appendix discusses approximating histories using subsets of the
principal components.

\section{Methods}
\label{sec:methods}
We first describe the
 N-body simulation plus semi-analytic modeling and hydrodynamical simulations used to create
the stellar mass histories, and then our use of PCA to characterize
these, as well as the halo mass histories, star formation rate histories and specific star
formation rate histories.  
We describe our use of mixture models and k-means
clustering in \S \ref{sec:splitsamp} itself as these are only used briefly.

\subsection{Simulations}
Our main simulation source was the Millennium database,
\citep{Lem06}, built upon the Millennium dark matter simulation \citep{Spr05},
a 500 $h^{-1}Mpc$ side box with $2160^3$ particles.  We used the
galaxy histories from
the semi-analytic model MR7 of \citet{Guo13}, with the
WMAP7 cosmology:
 $\Omega_m = 0.272$, $\Omega_\Lambda=0.728$, $h =
 0.704$, $\sigma_8=0.807$, $n=0.961$.   
\begin{table*}
\centering 
\begin{tabular}{l|c|c|c|c|c|c|c|c|c|c|}
$\frac{M_\star(z=0)}{10^{10}M_\odot}$
&median $\log_{10} \frac{M_h}{h^{-1}M_\odot}$&$
N_{\rm gal}$ & $f_{\rm sat}$&$\frac{\sum var(m) \times (10^{10} M_\odot)^2}{M_\star(z=0)^2}$&
$\frac{var(0)}{\sum var(m)}$&$frac_3$&``lowpc''/``highpc''&``lowk''/``highk'' \\ \hline
0.1 $\pm 2\%$  & 10.7&$\sim$43,000  
 &0.47
&1.1 & 0.77 &0.92  
& 0.35/0.32 & 0.29/0.20 \\ \hline
0.3 $\pm 2\%$  &11.3 &$\sim$11,000 
&0.46 
& 0.96 & 0.80 & 0.94 
&0.40/0.32 & 0.34/0.17  \\ \hline
1.0 $\pm 2\%$  &11.9 &$\sim$12,000 
  &0.42   
&0.95& 0.79 & 0.93 
&0.43/0.30 &0.40/0.18   \\ \hline
3.0 $\pm 2\%$  &12.1 &  $\sim$19,000
  &0.34  
&0.96&0.73&0.91  
& 0.40/0.29 & 0.32/0.21\\ \hline
10.0 $\pm 2\%$  &13.2 &$\sim$1300
& 0.22  
&1.15 & 0.65 &0.86  
& 0.23/0.32 &0.24/0.34   \\ \hline
20.0 $\pm 5\%$  & 14.0 &$\sim$300 
&0.10  
&0.90 &0.65  & 0.87 
& 0.23/0.31&0.41/0.40    \\
\hline
\hline
3.0 (+20\%,-17\%) &12.2 & 800&0.28
& 0.78 & 0.70  & 0.91 & 0.36/0.25 & 0.17/0.35  \\
(AGN+SN) & & & & & & & &  \\ \hline
3.0 (+20\%,-17\%)  &12.0 & 800 &0.25 
&0.42& 0.74  & 0.93  & 0.28/0.29 &0.27/0.52 \\  
(SN only) & & & & & & & &\\  
\hline
\end{tabular} 
\caption{The six Millennium (top) and two OWLS (bottom) samples,
with $z=0$ stellar mass range, median halo mass, number of
galaxies $N_{\rm gal}$, 
 satellite fraction, variance (in units of final stellar mass${}^2$),
fraction of
variance in $a_0$ and in $a_0,a_1,a_2\equiv frac_3$ combined (defined in Eq.~\ref{eq:mstarpc}), 
fraction of galaxies with
$\alpha_0$
(Eq.~\ref{eq:definealpha}) in the ranges
  [-1,-0.5),(0.5,1] (``lowpc''/''highpc''), and fraction of galaxies associated
  with
the lowest and highest k-means clustering paths
in \S \ref{sec:splitsamp} (``lowk''/''highk'').  See text.}  
\label{tab:samprops}
\end{table*}
We considered 6 different final ($z=0$) stellar mass bins,
$M_\star(z=0)/M_\odot\sim (10^9, 3\times
10^9, 10^{10}, 3 \times 10^{10}, 10^{11}, 2\times 10^{11})$. 
Several properties of the samples are
shown in Table \ref{tab:samprops} (e.g.\ the range of
final stellar masses, median halo mass, number of galaxies). 
The other columns are described and defined in subsequent sections.

For the three highest final stellar mass bins we took all such galaxies
in the simulation, for the three lower mass bins we instead took a large number
of galaxies with adjacent Peano Hilbert indices, i.e. in nearby
volumes within the simulation (the smallest volume used was
$\sim$1/4 of the simulation, large enough to not be affected by sample
variance).  The output times are equally spaced in
$\log_{10} a$; we took the 42 steps numbered 20 to 61 (starting at redshift 4.5 and
ending at redshift 0).  We also downloaded histories of halo virial
mass $M_{\rm halo}$ (equal to the infall mass for satellites, i.e. after a galaxy becomes a
satellite its halo
mass remains fixed at its infall mass), host halo mass $M_{\rm host}$
(which differs from virial
mass for satellites), star formation rates, history of centrality
(central or satellite), infall time (if satellite), g and r
colours for SDSS including dust, and stellar bulge mass.\footnote{We
  thank G. Lemson for help in navigating the database.}  For terminology,
a satellite galaxy is a galaxy that has fallen into a halo of a larger
galaxy (the infall time is when this happens, and the satellite
galaxy's halo is then called a subhalo).  
The Millennium simulation combines dark matter halo histories and a model of
how gas and subgrid properties behave as dark matter halos evolve to 
predict corresponding properties such as stellar mass, star formation, etc.

We also considered simulations from the OWLS project \citep{OWLS},
which include gas properties in the simulation itself.  We used galaxy
histories from two OWLS models: AGN+SN \citep[``AGN'', for details
about the AGN feedback model see][]{BooSch09} and SN only
\citep[``REF'', for details about the SN feedback model
see][]{DalSch08}. For more discussion of the these simulations and
their galaxy property results, see also
e.g. \citet{Sal10,vdV11,vDa11,Bry12,vdVSch12,Haa13a,Haa13b}.  These
simulations use a different cosmology than the Millennium samples
($\Omega_m = 0.238, \Omega_\Lambda=0.762, h = 0.73$, $\sigma_8=0.74$,
$n=0.951$), in particular a lower $\sigma_8$, implying slightly later
structure formation. They have $512^3$ dark matter and $512^3$
baryonic particles in a 100~comoving~$h^{-1}$Mpc side box.  We
concentrated on a single final stellar mass range, $\sim 3 \times
10^{10}M_\odot$, with 800 galaxies, at 20 redshifts from $z=0$ to 5
(with varying time steps, shown as the discrete points in
Fig.~\ref{fig:pc0comp}).  Higher stellar mass samples are too small
due to the simulation volume, and lower stellar mass samples cannot be
traced very far back in time due to resolution, so we limit ourselves
to only one final stellar mass bin.  These two OWLS samples can help
indicate trends which change with variations in gas physics or other 
assumptions.  This is especially useful given studies which contrast
properties between semi-analytic histories and histories inferred
from observational data sets (e.g., \citet{Wei12,MitLacColBau14};
\citet{Wei12} consider hydrodynamic simulations as well).

In both the Millennium and OWLS simulations, halos are identified by using a Friends-of- Friends (FoF)
algorithm.  If the separation between two dark matter particles is less
than 20 per cent of the average separation (the linking length b =
0.2), they are placed in the same group.  Subsequently \textsc{subfind}
\citep{Spr01,Dol09} is used on the FoF output to find the gravitationally bound
particles and to identify subhalos. For OWLS, baryonic particles are linked
to a halo or subhalo if their nearest dark matter neighbour is in that
halo,
 and galaxy stellar masses are calculated by adding up all the mass in stars
in the halo or subhalo to which the galaxy belongs.

\subsection{PCA decomposition of histories}
\label{sec:defpca}
We start by interpreting a galaxy's stellar mass $M_\star$ at a given
time $t_i$ as the galaxy's position along a $t_i$ coordinate axis, in
a space with as many dimensions as there are time steps $t_i$ (which
we choose to be the output times of the Millennium and OWLS models
respectively).  In this way, each galaxy history becomes a vector, with
each vector component corresponding to the galaxy's stellar mass at
a particular output time.
Every galaxy (vector) is then a single point in an $N$-dimensional space, where $N$ is the number of time outputs.  To apply
PCA, we shift the coordinates to make the average path
\begin{equation}
M_{\star,{\rm ave}} (t_i) = \frac{1}{N_{\rm gals}} \sum_{\rm gals  \;} M^{gal}(t_i)
\end{equation}
the origin, also a single point, in this space.  
All the galaxy histories then form a cloud of some shape around this
origin.\footnote{For each stellar mass sample, we also rescaled the
full stellar mass history of each galaxy so that its final stellar 
mass was equal to the median of the
sample.  This way we are focussing on fluctuations around the average,
not trends due to changing the final stellar mass.}

PCA characterizes the shape of this cloud, i.e.\ the
fluctuations.
So we are concerned with each history's 
deviation from the average,
$M_\star(t_i)-M_{\star,{\rm ave}} (t_i)$.  We can assign to each
galaxy
history a squared distance as well,
\begin{equation}
\Delta M^2= \sum_i  (M_\star(t_i)-M_{\star,{\rm ave}} (t_i))^2 \; .
\label{eq:distdef}
\end{equation}
Note that our choice of time output spacing affects $\Delta M^2$, as
each output time, or coordinate direction in this space, is given the
same weighting.  (However, as we describe below, we can often smoothly
interpolate results from one time output spacing to another.) 
The covariance matrix of the fluctuations from the average path
is given by
\begin{equation}
\begin{array}{l}
\frac{1}{N_{\rm gals}} \sum_{{\rm gals}} (M^{\rm gal}_\star(t_i)-M_{\rm ave} (t_i))
(M^{\rm gal}_\star(t_j)-M_{\rm ave} (t_j)) \\
=\langle (M_\star(t_i)-M_{\rm ave} (t_i)) (M_\star(t_j)-M_{\rm ave}
(t_j)) \rangle \; \\ 
\equiv C_{ij}
\end{array}
\end{equation}
This sum is straightforward to calculate from a set of histories and
results in an $N\times N$ matrix, with $N$ the number of time outputs.

Diagonalizing the covariance matrix to find basis fluctuations which
have zero covariance with each other is
the main idea behind PCA.  One reference for
principal component analysis is \citet{Jol02}, for
other applications to galaxies see
e.g., \citet{ConRydWei01,Mad02,Yip04a,Yip04b,Lu06,Bud09,Che12}, for cluster observables see,
e.g., \citet{NohCoh12} and for dark matter halos see
\citet{Jee11,SkiMac11,WonTay11}.  \citet{WonTay11} consider
halo histories and is most closely related to the work here.  PCA of
stellar population synthesis models and fixed time observables is considered in
\citet{Che12}.

The eigenvectors of the covariance matrix, $PC_n (t_i)\equiv PC_n$, 
are the
principal components, with eigenvalues $var(n)$ giving the 
contribution to the
total variance in the direction of $PC_n$.  Using the $PC_n$
as
basis axes (instead of the $t_i$) essentially rotates the
coordinate system in the (42-dimensional, for Millennium) space.  The
$PC_n$ can have any length, and arbitrary sign; we normalize
the $PC_n$ to length~1, and choose the sign convention that the
sum $\sum_{t_i} PC_n$ is positive.  We order the $PC_n$ so 
that $PC_0$ contributes the most to the total variance, and then $PC_1$, etc.,
i.e. for the associated variances, $var(0)>var(1)> var(2)$, and so on.  

Every galaxy's history $M_\star(t_i)$ corresponds to a set of $a_n$
coefficients ($n=0,1,2,...N-1$) of the $PC_n$,
\begin{equation}
M_\star(t_i) - M_{\star,{\rm ave}}(t_i) =
\sum_n a_n
PC_n  \times 10^{10} M_\odot\; .
\label{eq:mstarpc}
\end{equation}
We will choose $a_n$ and $PC_n$ to be dimensionless. 
These modes can be thought of roughly
as akin to Fourier modes, for fluctuations around an
average galaxy stellar mass history, with the $a_n$ similar to
Fourier coefficients.  However,  the shapes of the modes
(the $PC_n$) also contain information, as they are 
determined by the combined properties of the members of the sample whose fluctuations they describe.
\begin{figure}
\begin{center} 
\resizebox{4.3in}{!}{\includegraphics{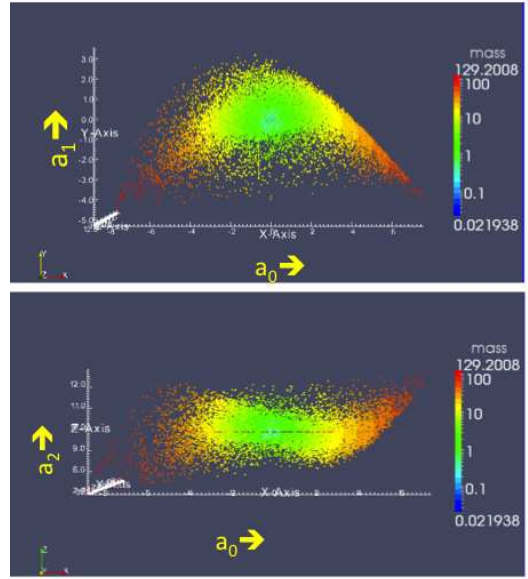}}
\end{center}
\caption{Each galaxy history is a point in the (42 or 20-dimensional) 
space spanned by $PC_n$.  For all galaxies in the
Millennium $M_\star(z=0)\sim 3 \times
10^{10} M_\odot$ sample, our fiducial sample, we show $PC_n$ components in the first three
dimensions (these dimensions have the largest spread in values).  We take
$x,y,z$ to correspond to the directions $PC_0,PC_1,PC_2$, so 
$a_0$ is shown along the $x$-axis, $a_1$ along the $y$-axis, 
and $a_2$ along the $z$-axis.
The $(a_0,a_1)$, i.e., $(x,y)$ plane is shown in the top panel, and the $(a_0,a_2)$, i.e., $(x,z)$ plane
in the bottom panel.  The colour (``mass'') indicates $\Delta M^2 =
\sum_{n=0}^{41} a_n^2 \times (10^{10} M_\odot)^2$, with average
8.6 $\times (10^{10} M_\odot)^2$ 
and median 4.6 $ \times (10^{10} M_\odot)^2$.  
The direction of $PC_0$ has the largest variance by
construction.  The distribution of
$a_0$ and of the signed fractional contribution to each galaxy's 
$\Delta M^2$, $\alpha_0$, are shown for this sample in
Fig.~\ref{fig:pc0props}.  As can be seen, the range of $a_0$ is much
larger than that of $a_1$ or $a_2$, i.e. $PC_0$ dominates, and
although the shape is not Gaussian in each direction, the fact that
the coefficients do tend to be near their central value is visible as well. 
}
\label{fig:3pcshape}
\end{figure}
Fig.~\ref{fig:3pcshape} shows $a_0,a_1,a_2$
for the $\sim 19,000$ stellar mass histories of Millennium sample
galaxies with $M_\star(z=0) \sim 3.0 \times 10^{10} M_\odot$.  This set of
galaxies will be our fiducial sample for examples below.  In
Fig.~\ref{fig:3pcshape}, the colour corresponds to the
full $\Delta M^2$: 
\begin{equation}
\Delta M^2 = \sum_{n=0}^{41} a_n^2 \times (10^{10} M_\odot)^2\; ,
\label{eq:distan}
\end{equation}
a rewriting of Eq.~\ref{eq:distdef}. 
That is, $a_0,a_1,a_2$ are directly visible, while the colour
includes the size of the remaining contribution from $a_3,...,a_{41}$.  The
median $\Delta M^2$ is $4.6 \times (10^{10} M_\odot)^2$, while its average  
is $\sum_n var(n) \times (10^{10} M_\odot)^2 =8.6\times (10^{10} M_\odot)^2$.  Its distribution has a long tail,
as seen by the ``mass'' range in
Fig.~\ref{fig:3pcshape}.\footnote{The large tail visible in the
  $\Delta M^2$ distribution is dominated by galaxies with very large
jumps in stellar mass near or at the final time.  Discarding or
keeping
the 20 or
fewer galaxies per sample with $\Delta M^2$ larger than 10 times the
sample variance changes the sample total variance, fraction of
variance due to $PC_0$, and correlations discussed later
by a few percent or (usually) less.  The leading principal component
$PC_0$ is essentially unchanged.  As the source of the stellar mass
jumps is unclear, and their effects minimal, we
  include these galaxies in our analysis below. }  
The average $\Delta M^2 / M_\star^2(z=0)$ is shown in Table
\ref{tab:samprops} for all samples.

The rewriting in Eq.~\ref{eq:mstarpc} holds for any change of basis,
not just the $PC_n$ chosen to diagonalize the covariance matrix.
However,
because the $PC_n$ are principal components,
the $a_n, a_m$ have zero covariance for $m \ne n$,
$\langle a_m a_n \rangle = \delta_{nm} \, var(n)$.    A quantity we will
also use later is 
the signed fraction of $\Delta M^2$ due to 
$PC_n$ for a given galaxy, i.e.
\begin{equation}
 \alpha_n \equiv sign(a_n) \frac{ a_n^2 \times (10^{10} M_\odot)^2}{\Delta M^2 }\;.
\label{eq:definealpha}
\end{equation}  
The $\alpha_n$ indicate
how much each particular $\pm PC_n$ contributes to a 
galaxy's $\Delta M^2$, its separation from the average history.
A galaxy with $\alpha_0 =1$ for instance, is proportional to $PC_0$,
but can have an arbitrary $\Delta M^2$ ($\propto a_0^2$) 
from the average history.

Tracking issues motivated us to modify the above procedure for 
the OWLS stellar mass histories.  The stellar masses of
Millennium galaxies do not decrease between time  
steps, however in the OWLS models,
75\% of the AGN+SN and 50\% of the SN only histories had stellar mass
loss at one or more step, and 49/800 (AGN+SN) and 18/800 (SN only) galaxies
lost over half of their mass between two consecutive steps.  
This latter huge mass
loss seems to be due the way stellar mass is assigned to
galaxies when a galaxy changes from a central to a satellite galaxy
and back again.  We dropped these histories with the most extreme mass
loss (i.e.\ 50 per cent or more) in computing and ordering the 
principal components, but then included all galaxies to calculate the
variance due to each principal component.  (This occasionally leads
the
fraction of variance due to a $PC_n$ to increase with $n$.)

For the Millennium samples, we also briefly compare PCA results for stellar mass histories
with their counterparts for galaxy halo histories,
star formation rate histories and specific star formation rate
histories.
We modified how we applied PCA for these histories in the following
ways and for the following reasons.  
A first change is that these other histories are not rescaled to be identical
at $z=0$.  The  
rescaling of stellar mass histories to get the same $z=0$ stellar mass
in each sample was to remove the effects of finite bin size in our
samples, i.e.
fluctuations due to 
slight changes in final stellar mass.  In contrast, 
for each essentially
fixed final stellar mass sample, the differences in final halo
masses, star formation rates and specific star formation rates are all
physical properties of interest.
A second change was the use of logarithms for halo mass.
As halo histories have a very large spread in final halo mass values,
those with large final mass heavily dominate the principal components.
This is somewhat improved by using $\log_{10} M_{\rm halo}$. (In part this
difficulty did not arise for stellar mass because we worked with
samples sharing essentially the same final stellar mass.   We could
have
used 
$\log_{10} M_\star$ for stellar mass histories as well, but chose not
to because it tends to emphasize the early
times of histories which are more sensitive to resolution
effects and less related to final time observables of interest to us.)
In addition, halo mass histories for galaxies which
switch from central to satellite and back again, or are more than
3 times the root mean square $\Delta (\log_{10}M_{\rm halo}/h^{-1}M_\odot)$ from
the median are excluded when
finding the halo bases
$PC_{n,{\rm halo}}$ in order for the components to represent most of the
histories
rather than the large remaining outliers. We include all haloes when calculating the $PC_{n,{\rm halo}}$ contributions to the
variance.  
This
leads to the non-monotonic behaviour with $PC_{n,{\rm halo}}$ seen for
some of the $n$ values in Fig.~\ref{fig:3pctypes}
below.  Another subtlety of halo histories is that, in simulations, the halo mass of a
satellite is fixed to be the halo mass right before it became a
satellite. These constant satellite halo masses and the fact that
some galaxies switch from central to satellite and back several times tend to complicate
interpretation.  As a result,
while we will mention some results below, we will not consider these
histories in as much depth as the stellar mass histories.
Note that \citet{WonTay11} considered halos rather
than galaxy halos and thus did not encounter these issues, which
are associated with subhalos, not halos.

\section{PCA Classification}
\label{sec:pca}
We now discuss the properties of the PCA classification of the
galaxy stellar mass histories.  As PCA concerns fluctuations around
the
origin in a 42 or 20-dimensional space for our samples, for each
sample we start by
showing the origin
(the average stellar mass history).  We then turn to the shape of
each sample's distribution around the origin ($var(n)$ of each of the
$PC_n$), touching
 upon the $PC_n$ contributions for the halo mass, star formation rate and
specific star formation rate counterparts.
We comment briefly on the accuracy of approximating galaxy histories
with a certain number of $PC_n$ (covered in more depth in the Appendix), and the properties of the $PC_n$, and
$PC_0$ in particular, as functions of time.
\begin{figure}  
\begin{center} 
\resizebox{3.5in}{!}{\includegraphics{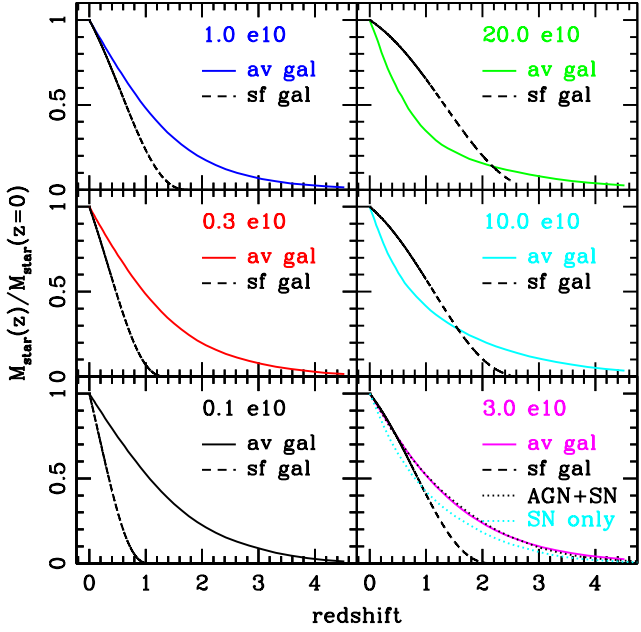}}
\end{center}
\caption{
Average histories for the 8 final stellar mass samples, rescaled to
final stellar mass of each sample (solid lines for Millennium, dotted
for OWLS) in order to highlight similarities and differences between them.
Shown as well (dashed lines) is the
history of a galaxy with the same final mass on the
observationally inferred
  star-forming main sequence from \citet{Lei12}.  For high final
  stellar
mass galaxies, the average stellar mass growth is often higher at late
times than
that of a galaxy on the star forming main sequence. 
}
\label{fig:avhist}
\end{figure}

The average histories for all samples are shown in
Fig.~\ref{fig:avhist}, rescaled by their final stellar mass to allow
easier intercomparison.  These curves are each the origins in their
respective (42 or 20-dimensional) spaces spanned by their PCA $PC_n$.
Also shown, for reference, are the corresponding stellar mass
histories of galaxies on the star forming main sequence with the same
final stellar mass, inferred from observational data and using
different
assumptions than the simulations,  by \citet{Lei12}
(amended\footnote{We thank S. Leitner for sending us the
  correction.} Eq.~A3, shown in his Fig.~9).  
The average histories for the
two higher final stellar mass samples grow more quickly at late times
than a galaxy on the star forming main sequence, presumably due to
higher than main sequence star formation rates, stellar mass gain
through mergers, or some difference between the Millennium models and
the assumptions for the inferred histories from observation.  These two higher final stellar mass bins also
exhibit other differences in their histories from those with lower
stellar mass, as will be seen later.

\begin{figure}
\begin{center} 
\resizebox{3.5in}{!}{\includegraphics{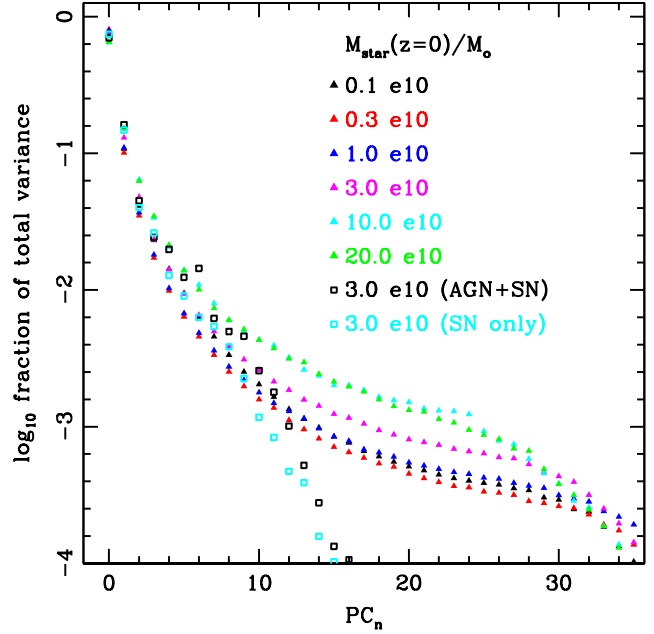}}
\end{center}
\caption{Logarithm of the fraction of the total variance $var(n)/\sum_m var(m)$ 
due to $PC_0$, $PC_1$, up to $PC_{35}$,
for the 6 Millennium samples, ($M_\star(z=0)/M_\odot \sim $
$10^9, 3 \times 10^9, 10^{10}, 3\times 10^{10}, 10^{11}, 2\times 10^{11}$), and the two OWLS
samples ($M_\star(z=0)/M_\odot \sim 3 \times 10^{10}$), with colours as in Fig.~\ref{fig:avhist}.
There are fewer outputs for the OWLS
runs, hence fewer $PC_n$.  Contributions to the variance
around the average histories (shown in Fig.~\ref{fig:avhist})
 quickly drop to below
a percent after the first few $PC_n$. 
}
\label{fig:pcfrac}
\end{figure}

The fraction of the variance $var(n)$ due to each $PC_n$ characterizes
the ensemble of stellar mass histories.
As seen in Table \ref{tab:samprops} and Fig.~\ref{fig:pcfrac}, 
the leading three $PC_n$ contribute a significant fraction, $\sim 90\%$, of
the total variance around the average history, with $PC_0$ strongly
dominating (note that the y-axis is logarithmic).\footnote{Truncating the earlier time steps to only go
back to redshift 1.7 instead of 4.5 for the Millennium sample gives
slightly stronger dominance of $PC_0$, and moves the peak position 
of $PC_0$ slightly, the inner product of the truncated and full
$PC_0$ is 1.00.} 
For the Millennium histories this
means the remaining $\lesssim10\%$ of the variance is split among the
remaining 39 $PC_n$, for the OWLS histories this fraction of
the variance is shared amongst the remaining
17 $PC_n$.
The predominance of
$PC_0$ is also indicated in Fig.~\ref{fig:3pcshape} by the much larger
range of values for the $a_0$ distribution, relative to $a_1$ and $a_2$.

\begin{figure}
\begin{center} 
\resizebox{3.5in}{!}{\includegraphics{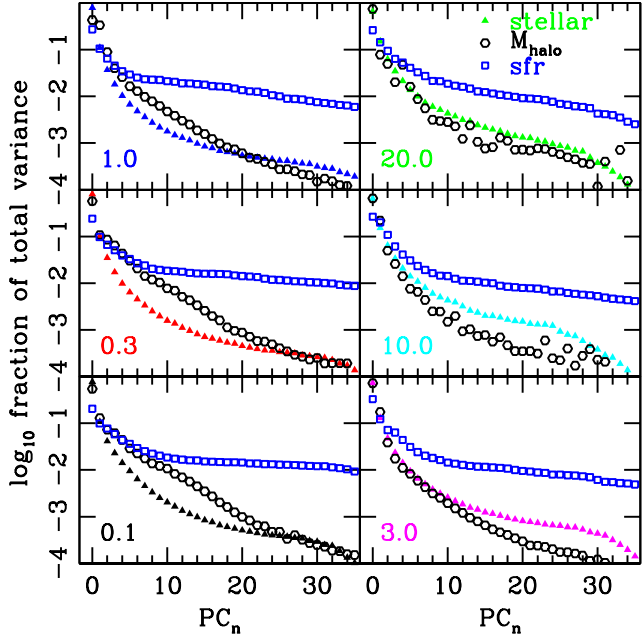}}
\end{center}
\caption{Logarithm of the fractional variance due to each corresponding PCA
  component for stellar mass (solid triangles), $\log_{10} M_{\rm halo}$ 
  (open hexagons, black) and star formation
rate (open squares, blue) histories, for all 6 Millennium samples.  The stellar mass histories
are the same as in Fig.~\ref{fig:pcfrac}, they and the
  halo mass histories are dominated by the first few components, while
  those for star formation rate require a large number of components
to describe the same fraction of total variance around the average
history.  $M_\star(z=0)/10^{10} M_\odot$ is at lower left in each panel.}
\label{fig:3pctypes}  
\end{figure} 
For some context and comparison, Fig.~\ref{fig:3pctypes} shows
the principal component contributions for the Millennium stellar mass histories
of
Fig.~\ref{fig:pcfrac} again, but now alongside their counterparts for
the star formation rate and $\log_{10} M_{\rm halo}$ 
(as described earlier, recall $\log_{10} M_{\rm halo}$ is used due to the large halo mass
range) for the same samples. 
The star formation rate histories are not dominated by the
first few principal components:  on average less than half of the total
variance is due to the first three $PC_{n,{\rm sfr}}$.  In contrast,
just as for the 
stellar mass histories, the halo mass histories are strongly dominated
by the first few principal components. On average 87 per cent of the
total variance is due to the first three $PC_{n,{\rm halo}}$. This
fraction of the variance changes noticeably with
sample, perhaps in part due to differences in their satellite
fractions
(Table \ref{tab:samprops}).  Again, the scatter up and down for $PC_{n,{\rm
    halo}}$, especially apparent in the highest mass samples, happens
because the $PC_{n,{\rm halo}}$ are calculated (and assigned an order) 
with a subset of the
galaxies, as described at the end of \S \ref{sec:defpca}, while the
full
set of galaxies is used to compute their variance contributions.

We also considered
the specific star formation rate (not shown for brevity).
Its principal component fractional contributions to the variance fall between those for
 star
formation rate and stellar mass in Fig.~\ref{fig:3pctypes}, with the first
three components contributing on average 3/4 of the variance.
For higher $n$ components, the specific star formation rate
contributions are close to those for $\log_{10} M_{\rm halo}$ for the three
lowest final stellar mass samples ($M_\star(z=0)<3 \times
10^{10}M_\odot$).  For $M_\star (z=0) \geq 3 \times 10^{10}
M_\odot$, the higher $n$ specific star formation rate principal components
contributions tended to be much higher than for $\log_{10}$ halo
mass. 
For all final stellar mass samples,
the fraction of the variance due to the $n$th
principal component for specific star formation rate
drops below that for stellar mass around $n=20-22$.

In addition to studying contributions to the total scatter of the
sample, the $var(n)$, one can also ask about contributions to
individual galaxy histories, in particular how well keeping only a small
number of $PC_n$ modes in the expansion shown in Eq.~\ref{eq:mstarpc}
approximates individual galaxy stellar mass histories.  Once a galaxy
stellar mass history is approximated by about 20 $PC_n$, 95\% of
galaxies are within $< 0.15 M_\star(z=0)$ of their true history, 
and within 10\% of
their true $\Delta M^2$.  See the Appendix for 
figures showing more detailed trends for how these approximations work as
final stellar mass and number of $PC_n$ are varied. 

\begin{figure}
\begin{center} 
\resizebox{3.5in}{!}{\includegraphics{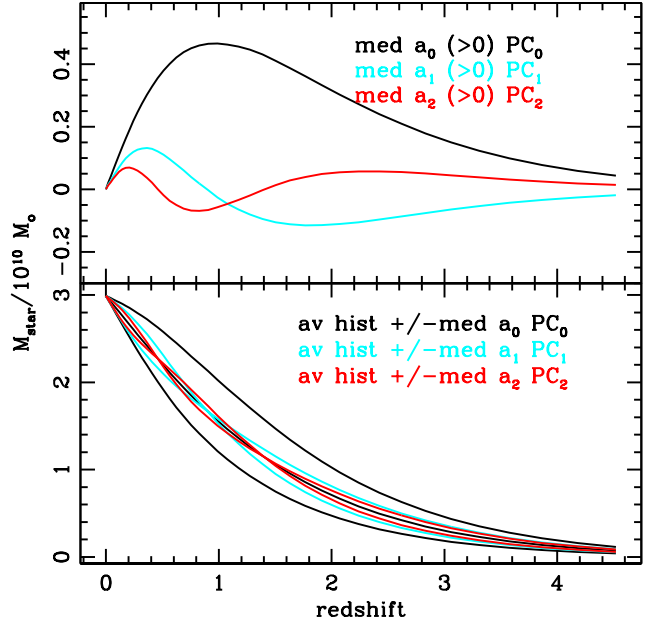}}
\end{center}
\caption{
Dominant paths contributing to 
fluctuations around the average
galaxy stellar mass history, for galaxies in the fiducial
(Millennium $M_\star(z=0) = 3\times 10^{10}M_\odot$) sample.
Top:  the three most dominant PC's, multiplied by their
respective median positive coefficients $a_0,a_1,a_2$.   
Bottom:  the fluctuations added to the average history, with
positive and negative coefficients set to their respective median values for
the fiducial sample.
}
\label{fig:first3}
\end{figure}

We now turn to the $PC_n$ themselves as functions of redshift.  
The leading three $PC_n$ for the fiducial sample (where they
account for 91\% of the variance) 
are shown in Fig.~\ref{fig:first3}.  Again, this is the Millennium $M_\star(z=0)\sim 3 \times
10^{10} M_\odot$ sample.  The coefficients of each
are set to their median positive (for $+PC_n$) or
 negative (for $-PC_n$) values.
The $n$th $PC_n$ crosses zero $n$ times.
For
the highest $PC_n$, there is a lot of 
``ringing'' at the highest redshifts, with large amplitudes (but with
small coefficients $a_n$).  
The extrema 
are spaced more regularly in lookback time than in redshift.
Regularity was also seen
by \citet{WonTay11} when using the scale factor $a$ as time coordinate for their halo history principal components.
They mentioned the similarity to Fourier modes, although unlike
Fourier
decomposition, with the
PCA expansion both the coefficients and the basis components (i.e. basis modes) are determined by the
sample.\footnote{
Because the fractional variations in log$_{10} M_{\rm halo}$ and specific star
formation rate are larger at high redshift, as compared to stellar
mass and star formation rate, the largest variations in $PC_{0,{\rm halo}}$
and $PC_{0,{\rm ssfr}}$ occur at
significantly higher redshift than for $PC_0$ and $PC_{0,{\rm sfr}}$, respectively.}

\begin{figure}
\begin{center} 
\resizebox{3.5in}{!}{\includegraphics{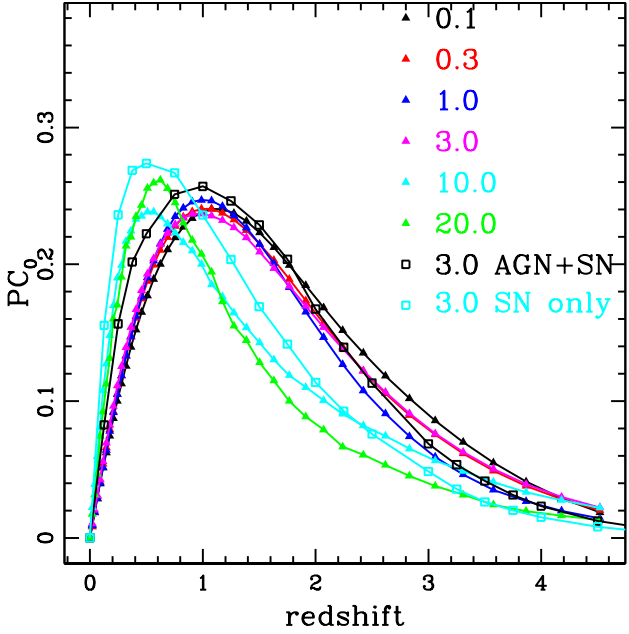}}
\end{center}
\caption{
The normalized  (for Millennium) leading fluctuation around the average galaxy stellar 
mass history,
$PC_0$, for all 8 samples, colours as in
Fig.~\ref{fig:pcfrac} (listed in terms of $M_\star(z=0)/10^{10}
M_\odot$ as well).   The two OWLS $PC_0$ are normalized to length
$\sqrt{20/42}$ to compensate for the smaller number of output times.
The $PC_0$ for the four lowest $M_\star(z=0)$ Millennium samples and the AGN+SN OWLS
samples have
inner product
$>0.99$ with each other, similarly for the two highest
$M_\star(z=0)$ 
Millennium samples and the SN only OWLS sample. 
When one $PC_0$ is taken from both groups
the overlap is smaller, dropping to 0.91.  In the Millennium sample,
there seems to be a qualitative change above $M_\star(z=0)=3\times
10^{10}M_\odot$.  The triangles and squares show the actual output
times of the simulations (which imply a certain weighting of different
periods of time for the PCA analysis).
}
\label{fig:pc0comp}
\end{figure}
The leading component, $PC_0$, dominates the variance,
and as will be seen below, many of the individual galaxy histories as
well.   
Fig.~\ref{fig:pc0comp} shows $PC_0$ as a function of redshift for all
8 samples.  
The data points are the simulation output redshifts, whose spacing
provides an implicit
weighting for $\Delta M^2$.  To allow easier visual comparison 
between $PC_0$ for 
the OWLS and Millennium models, we multiplied 
$PC_0$ for OWLS by $\sqrt{20/42}$, to compensate for its smaller
number (20) of output times compared to Millennium (42).  

We see that although 
$PC_0$ is dominant and has a single peak in all cases, 
differences appear as final stellar mass and simulation are varied.
The smoothness of the leading $PC_n$, in particular $PC_0$, allows
us to use linear interpolation to intercompare them between
simulations
with different numbers of output times.
$PC_0$ is similar
amongst the four lowest $M_\star(z=0)$ Millennium
samples and the AGN+SN OWLS sample, as well as amongst the two higher $M_\star(z=0)$ samples and the
SN only OWLS sample. Within these groups
inner products are $>0.995$, but between the two groups they can drop
to 0.91.\footnote{
 Note that the $PC_n$ vectors for different samples are fluctuations
  around different average histories.
The next leading $PC_n, n=1,2$ are also very close within the lowest
four and within the two
  highest $M_\star(z=0)$ Millennium samples.  For these the AGN+SN
  OWLS sample again has much higher overlap with the lower four, while the
SN only OWLS sample has larger overlap with the higher two.}  
The notable feature in
$PC_0$ is its peak, which is at lower redshift for the two highest
$M_\star(z=0)$ Millennium samples and the SN only OWLS sample,
but at approximately the same redshift for the lower 
$M_\star(z=0)$ samples and the AGN+SN OWLS sample.  
The
two OWLS samples share the same final stellar mass as the fiducial
sample, one of
the lower $M_\star(z=0)$ samples.  
The difference between the two OWLS samples is probably due to
overcooling in the SN only case, leading to too much late time star
formation as compared to observations.  (The source of the difference
between higher and lower final stellar masses is less clear, but
possibly related to the increased importance of mergers for the former.)  Thus we see how 
the behaviour of $PC_0$ discriminates between families of
histories with different physics and (sometimes) with different final
stellar mass.

In summary, $PC_0$ dominates the stellar mass histories for all the different samples, and the
first three $PC_n$ capture $\sim$90\% of the variance of the full sample.
In addition (see Appendix for further discussion),
a large fraction of individual galaxy stellar mass histories
can be well approximated by a subset of the $PC_n$.  The leading
$PC_0$ has one feature, a peak, and is similar for the
$M_\star(z=0) \leq 3 \times 10^{10}M_\odot$ Millennium samples and the
AGN+SN OWLS sample. A different $PC_0$ seems to be shared between
the two higher $M_\star(z=0)$ Millennium
samples and the SN only OWLS sample, peaking at a lower redshift.
The halo history (which is somewhat more difficult to interpret) PCA shows similar dominance of the leading few components,
while that for the specific star formation rate histories is slightly weaker.
Star formation rate histories require many more principal components to get a
similar
fraction of the variance compared to the stellar mass case. 

\section{PCA classification and other galaxy properties}
\label{sec:corrlns}

The PCA decomposition revealed that a few parameters characterize
a large fraction of the total deviation from the average stellar mass
history for the ensemble of histories, and that a large fraction of 
each galaxy's history is captured by the leading $PC_n$.
 We now look for relations of the PCA characterizations with other galaxy
properties, in particular their halo histories, specific halo and
stellar mass history times, and observables at a fixed
time (in our case $z=0$). 

We start with $\alpha_0$, the fraction of a galaxy's history due
to $\pm PC_0$, where $\alpha_0>0$ corresponds to a $PC_0$ 
contribution with
earlier stellar mass gain than the average history, and $\alpha_0<0$
to later stellar mass gain than the average history.

One interesting history to consider in terms of the leading fluctuation, $PC_0$, is the
observationally derived history of
a galaxy on the star forming main sequence (\citet{Lei12}, shown in
Fig.~\ref{fig:avhist}), for a given final stellar mass. 
We can write this history
in terms of $a_n$ for each sample, and calculate its corresponding
$\alpha_0$, which additionally quantifies the fractional contribution
of $PC_0$ to any history's $\Delta M^2$.
We find $\alpha_0= $ 
(-0.94,-0.97, -0.89, -0.54, +0.55, +0.89) for  Millennium $M_\star
(z=0)=$(0.1, 0.3, 1.0, 3.0, 10.0, 20.0 $\times 10^{10} M_\odot$) and 
(-0.54, -0.02) for OWLS (AGN+SN, SN only, both with $M_\star(z=0)= 3.0\times 10^{10}
M_\odot$).  That is, the star forming main sequence corresponds to
$\alpha<0$, late stellar mass gain, 
for the lowest final stellar mass histories in Millennium
and the AGN+SN OWLS sample.  This special history is in fact 
predominantly (contributing more than 90 percent of $\Delta M^2$) along the direction
of $-PC_0$ for the two lowest final stellar mass histories in
Millennium.  For the two highest final stellar mass
histories in Millennium,
the \citet{Lei12} observationally derived star forming main sequence has slower stellar mass growth than
the average history, and actually corresponds to 
early stellar mass gain in terms of $PC_0$ ($\alpha_0 > 0$).  It may
be that for these
galaxies late stellar mass gain is due to
mergers (which are more common for their likely hosts, halos with
large masses, which also tend to have less star formation),
starbursts, or perhaps some difference between the theoretical
modeling and the inferences from observations.

Considering other galaxy stellar mass histories,
one might guess that early stellar mass gain relative to
the average history, $\alpha_0>0$,
corresponds to a quiescent galaxy, for instance a quenched
satellite. (Quenched central galaxies may be expected to have relatively more late 
stellar mass gain from mergers.)  
Consistent with this,
for the lower stellar mass Millennium samples, 
satellites are an increasing fraction of the galaxies as $\alpha_0$
increases from negative to positive.
Similarly, $\alpha_0<0$, i.e.\ late stellar mass gain relative to the average history, is more common for low
final stellar mass centrals and this trend reverses as
$M_\star(z=0)$ increases. 
The interpretation is less clear for
the two higher final stellar mass samples, which have $\alpha_0 >0$
for the star forming main sequence, and
satellites which are either a constant or decreasing
fraction of the galaxies as $\alpha_0$ increases.

\begin{figure}
\begin{center} 
\resizebox{3.5in}{!}{\includegraphics{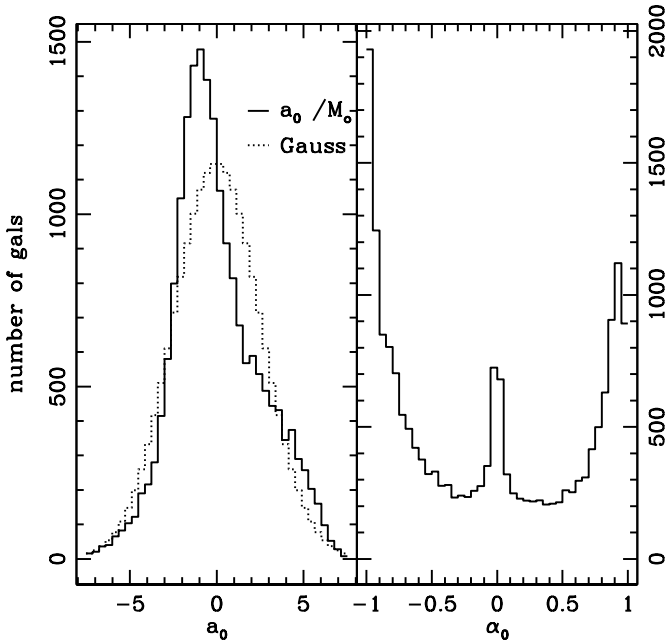}}
\end{center}
\caption{Left: distribution of $a_0$, the coefficient of $PC_0$ in a
galaxy's history, for
  the
fiducial (Millennium $M_\star(z=0) = 3\times 10^{10}M_\odot$) sample.   The distribution of $a_0$ is asymmetric
and narrower than a Gaussian (dotted line) with the same $\sigma (=2.5)$ and
normalization.
 Right: distribution of the signed fraction of each individual galaxy's total
variance $\Delta M^2$ due to $PC_0$, $\alpha_0$, which is positive or negative depending upon whether the
galaxy has $\pm PC_0$ in its history, again for the fiducial sample.
An $\alpha_0=1$ galaxy has all of its deviation from the average history
made up of $+PC_0$.  
}
\label{fig:pc0props}
\end{figure}

The fractions of galaxies with $\alpha_0<-0.5$ and $\alpha_0>0.5$,
 respectively ``lowpc''/``highpc'', are listed for all 8 samples in Table
 \ref{tab:samprops}.  These are galaxies where at least half of their
 $\Delta M^2$ is due to $PC_0$, i.e. to the simple shaped fluctuation
in Fig.~\ref{fig:pc0comp} determined separately for each sample.
For a multivariate Gaussian with the same variances as our
   fiducial model there would be 2/5, 1/5, 2/5 of the sample in the regimes
   $\alpha_0 \in$ [-1,0.5),[-0.5,0.5],(0.5,1].  In comparison, for
our samples
   there are relatively more galaxies in the central regions; in addition,
   the high and low $\alpha_0$ regions are not symmetric. 
The $\alpha_0$ distribution for all galaxies in
 the fiducial sample  is
 shown in the right panel of Fig.~\ref{fig:pc0props}, where peaks at
$\alpha_0 = \pm 1, 0$ are visible, characteristic of a dominant $PC_0$
in a many dimensional system.

Because halo mass history variances are also often dominated by
$PC_{0,{\rm halo}}$ (in fact slightly more so than the stellar mass
history variances by $PC_0$, for the three highest final stellar mass
samples), one might expect that many galaxies have sizable
contributions from $PC_{0,{\rm halo}}$ in their individual histories.
In terms of the analogue of
$\alpha_0$ for halo mass histories, $\alpha_{0,{\rm halo}}$, defined
via
$a_{0,{\rm halo}}$ as in Eq.~\ref{eq:definealpha}, we found
$|\alpha_{0,{\rm halo}}|> 0.5$ for
(52\%, 56\%, 24\%, 68\%, 69\%, 55\%) of the Millennium galaxies.
(This can be compared to the stellar mass history counterpart, the 
sums of ``highpc'' and ``lowpc'', in
Table \ref{tab:samprops}.) 
That is, for all but one final stellar mass sample, 
a large fraction of galaxies have more than half of their $ (\Delta \log_{10} M_{\rm halo})^2$
 along the $PC_{0,{\rm halo}}$ direction.  The $M_\star(z=0) =
 10^{11}M_\odot$ sample is likely an outlier because its $PC_{0,{\rm
     halo}}$ direction is associated with a low fraction, 42\%, of the
 total variance, as compared to the average fraction for the other samples
 ($\sim 2/3$).
In addition, $\alpha_{0,{\rm halo}}$ 
is also often strongly correlated
with $\alpha_0$, with
correlation coefficients (73\%, 68\%, 65\%, 64\%, 19\%,
54\%).\footnote{It is not clear why one of the high final
stellar mass samples gives a very low correlation.}
In the OWLS models the
correlations between $\alpha_0$ and $\alpha_{0,{\rm halo}}$ 
was roughly 1/3 in
both cases;
stellar mass history and halo mass history seemed less
correlated in the two hydrodynamical simulations by this
measure.
 This quantity would be interesting to
compare
in other simulated samples, although as noted earlier, issues of how halo mass
is defined for a galaxy when it becomes a satellite
 (fractions are shown for our samples in Table~\ref{tab:samprops}) and
satellite to central switches generally
 make interpretation of the halo histories more challenging.
 Relations between average halo mass and
stellar
mass and/or (specific) star formation rate\footnote{The 
star formation rate and specific star formation rate histories considered above for the Millennium samples are not as strongly
dominated by the first principal component, making the relevance of
the
first component's contribution less clear.  For the OWLS models,
the importance of the leading star formation rate and specific star
formation rate principal component is difficult to
estimate due to resolution issues.}
 have been derived observationally at fixed redshift \citep[e.g.][]{ValOst06,ConWecKra06,Mos10,Lea11,Red13} and relations have
also been found on average between stellar mass
and halo mass growth 
\citep{Mor09,YanMovdB09,Yan12,BehWecCon13a,BehWecCon13b,Hud13,MosNaaWhi13,Tin13,Wan13}.

\subsection{Correlations 
with history times}
\label{sec:corrhist}
Going beyond correlations with $\alpha_{0,{\rm halo}}$,
we searched for further relations between stellar mass PCA quantities 
and theoretical and observational properties.
In principle such relations might differ
between simulations, however, as many quantities in the
OWLS models were not defined in the same way, we focus on the
Millennium samples only.
The PCA description quantities we considered were $a_0$, $\alpha_0$,
$\alpha_1$, $\alpha_2$, $\Delta M^2$ and $\Delta M_+^2 = \sum_{m=3}^{N}
a_m^2 \times (10^{10} M_\odot)^2$.  We will show mostly results for $a_0$ below, others will be
mentioned for the cases where we found large correlations.
The $a_0$ for a galaxy is the total amount of its history due to
$PC_0$, rather than the signed fractional contribution to $\Delta M^2$
measured by $\alpha_0$.  We use $a_0$ for these comparisons because
dependence upon $\alpha_0$ has
 much more structure due to the
three peak distribution of $\alpha_0$, visible in
Fig.~\ref{fig:pc0props}.   (Often the
correlation coefficients for $a_0$ and $\alpha_0$
are very similar, some examples are given of this below.) 

We first consider the relation between $a_0$
and several times which characterize
galaxy stellar mass or halo mass histories.
Unlike the $PC_n$ coefficients, these times do not say anything about the
shape
of the stellar mass history or the halo mass history aside from the time it reaches a certain
value.
For stellar mass history times, we take the redshifts $z_{10},z_{50},z_{90}$ where
a galaxy reaches 10\%, 50\% or 90\% of its final stellar mass (if the redshift
is before $\sim$ 4.8, the earliest redshift we have, we set it to 5).
For halo history we take times which have
been used to identify which galaxies might be quiescent in
the ``age matching'' model of \citet{HeaWat13}.  These are:\newline
$z_{12}$: the redshift at which
the galaxy's halo mass\footnote{Millennium halo masses are based upon $M_{200c}$, we use the estimate
$M_{\rm vir}$ = 1.22 $M_{200c}$ \citep[Fig.~1 in][]{Whi01}.} 
 $M_{vir}$ reaches $10^{12} h^{-1}  M_\odot$. \newline
$z_{\rm grow}$: the redshift when a galaxy's halo growth falls below a
certain rate.  We take the definition of \citet{Wec02} as suggested by \citet{HeaWat13,Hea13b},
$d\log M_{\rm halo}/d\log a\equiv \Delta\log M_{\rm FoF}/\Delta\log a
<2$. \newline
$z_{\rm infall}$: The redshift when a galaxy falls into another halo and becomes
a satellite.  \newline
$z_{\rm starve}$: The earliest redshift of $z_{12},z_{\rm
    grow},z_{\rm infall}$.\newline
Subhalo times are used if the galaxy is a
  satellite.
Two other previously discussed PCA history quantities are also included, $\Delta
M^2/M_\star(z=0)^2$, the total squared deviation of a particular
galaxy's history from the average, and $\alpha_0$, the fractional
contribution
to a galaxy's $\Delta M^2$ due to $PC_0$.

\begin{figure}
\begin{center} 
\resizebox{3.5in}{!}{\includegraphics{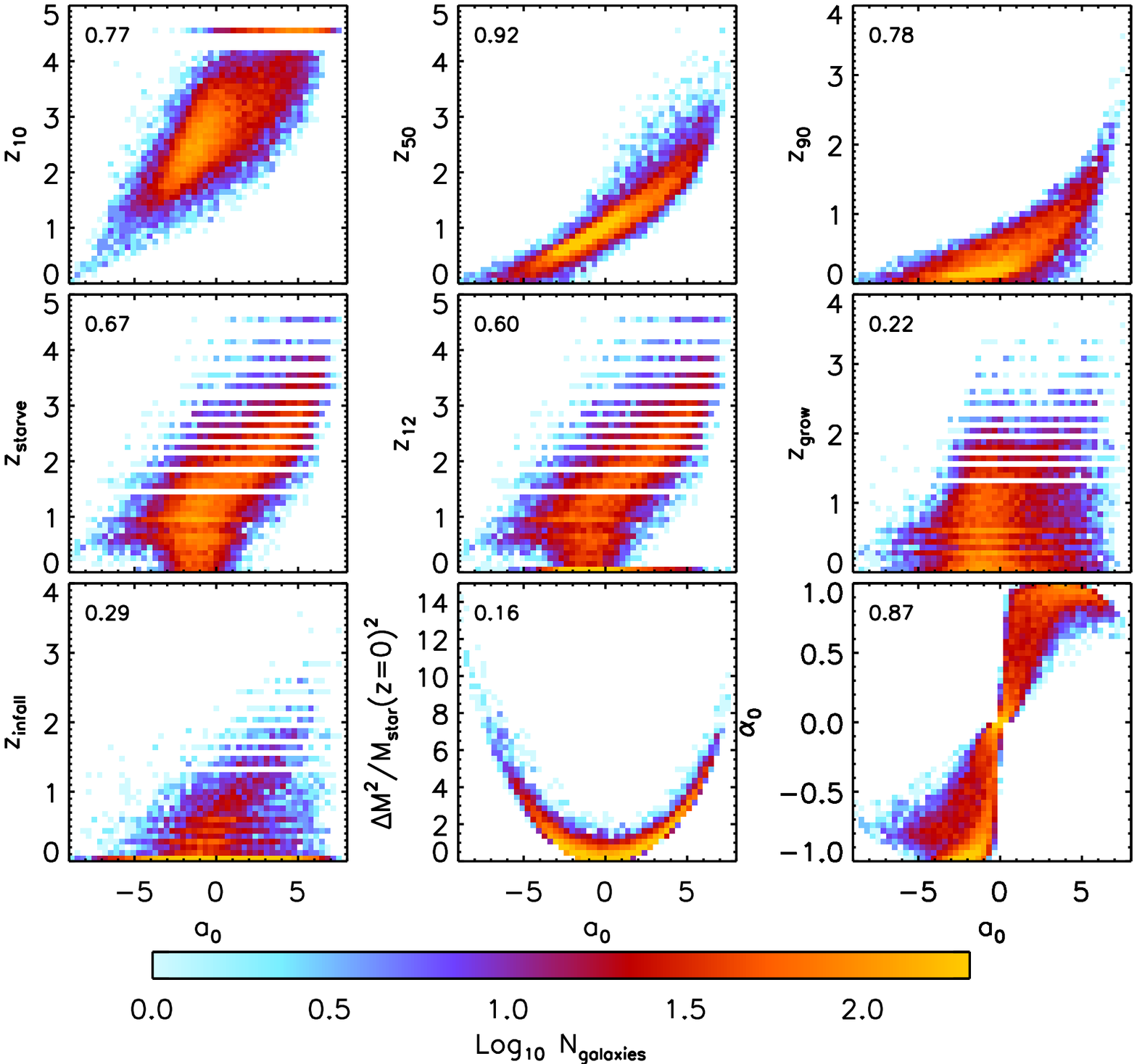}}
\resizebox{3.5in}{!}{\includegraphics{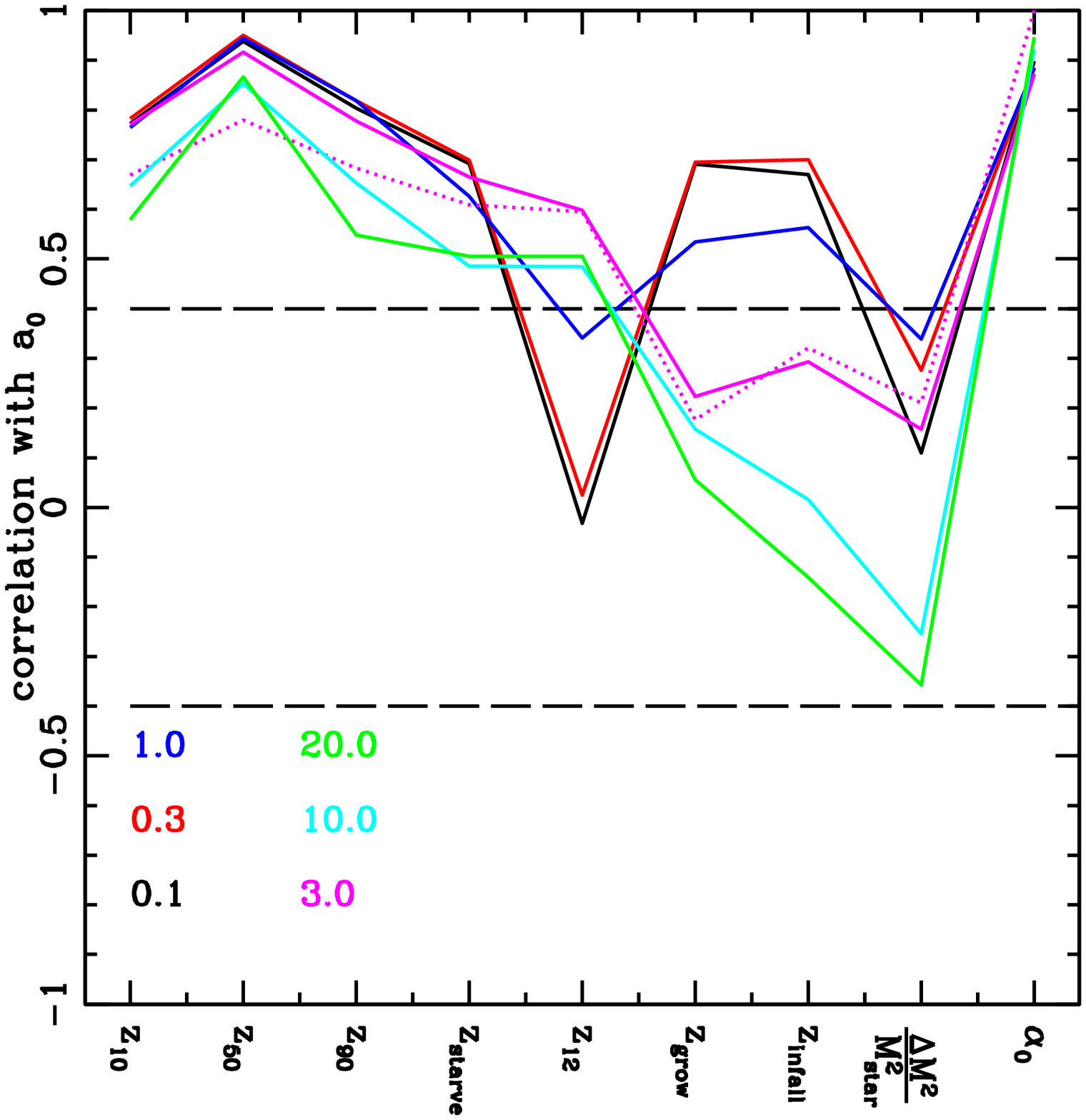}}
\end{center}
\caption{Comparison of $a_0$ with other galaxy history characterizations.
Top:  For the fiducial sample, distributions of the leading PCA
  coefficient, $a_0$, versus several halo history and stellar mass history
quantities.
Bottom:
The correlation coefficients for the above scatter plots and their
counterparts for all 6 Millennium samples.
Coloured curves are for different
final stellar mass samples, as indicated in the legend in terms of
$M_\star(z=0)/10^{10}M_\odot$ in the lower left corner.
The dotted line is the correlation coefficient for
$\alpha_0$, rather than $a_0$, i.e. the signed fractional contribution to
$\Delta M^2$ from $PC_0$, for the fiducial sample.
The times $z_{10}$, $z_{50}$, and $z_{90}$ are when galaxies
have reached 10\%, 50\% and 90\% of their final stellar mass, while
the halo history times $z_{\rm starve}$, $z_{12}$, $z_{\rm
  grow}$ and $z_{\rm infall}$ have been suggested as
determining the ``age'' of a galaxy.  See text for definitions. 
The measure $\Delta M^2$  is
total ``distance'' squared of a galaxy's stellar mass history from the
average, and $\alpha_0$ is the signed fraction of a galaxy's $\Delta M^2$ due
to
$PC_0$. Strong correlations are present for many of the characteristic
redshifts. 
}
\label{fig:histcorr}
\end{figure}

The distributions of these quantities relative to $a_0$ are shown at
the
top panel of Fig.~\ref{fig:histcorr}, for the fiducial sample.
Correlation
coefficients are shown in each panel in the upper left corner. In the
bottom panel we show the correlation coefficients again, along with their counterparts for the other 5 Millennium
samples, to illustrate trends with changing final stellar mass. 
Lines are drawn at $\pm 40\%$ to draw attention to large positive and
negative correlations.

Many of the correlations are large.
The large correlation of $a_0$ with the times $z_{10}, z_{50},z_{90}$ is not
surprising, as $a_0$ corresponds to 
a component of a galaxy's history with early 
(for a positive sign) or late (for a negative
sign) stellar mass gain, and $PC_0$
dominates the history of many of the galaxies  (Fig.~\ref{fig:pc0props}).\footnote{We also found, not
shown, that $a_1$ had a large correlation with $z_{10}$ for all
samples,
increasing with increasing final stellar mass (for the two highest
final stellar masses $a_2$ also had a correlation of 30\% or higher with
$z_{10}$,
and for the highest final stellar mass $a_1$ also had a large
correlation with $z_{90}$).  Note that these last two final stellar
mass 
samples had weaker dominance of $PC_0$ as well.}
The PCA description goes beyond these specific times in a galaxy's stellar
mass history to
say how much of that history's departure
from the average is due to a particular shape, i.e.\ due to $PC_0$.
A strong correlation with $z_{starve}$ is also not unexpected, as
earlier $z_{starve}$ means earlier shut off of star formation, which
again
(given that all galaxies in each subsample have the same final stellar
mass) leads to the expectation of large positive $a_0$ (early stellar
mass gain).  
The higher final stellar mass samples often have weaker correlations
of
these times
with $a_0$, which is likely a combination of $a_0$ not being as
dominant (for all measurements)
and star formation perhaps not being as major a component of stellar
mass
gain.  Some other differences
for different final stellar mass samples are easy to interpret,
e.g. the lower final stellar mass samples rarely reach halo 
masses large enough to have $z_{12} \ne 0$.  

In terms of other PCA related quantities,
the correlations with $\alpha_0$ are similar, 
as an example the fiducial sample's $\alpha_0$ correlations are shown as the dotted curve
in Figs.~\ref{fig:histcorr} and \ref{fig:obscorr}.  
Correlations with 
$\alpha_1, \alpha_2$ are
relatively small, except for the subsamples with $\alpha_1>0.5$
or
$\alpha_2 >0.5$ respectively (not shown). For these the large correlations which
appeared were strongly $M_\star(z=0)$ dependent. 
As can be seen, as $a_0$ increases, the scatter $\Delta M^2$ from the 
average history tends to 
increase for the four lower final stellar mass samples, and decrease
for the two highest.  

\subsection{Correlations 
with final time observables}
\label{sec:corrfinal}

\begin{figure}
\begin{center} 
\resizebox{3.5in}{!}{\includegraphics{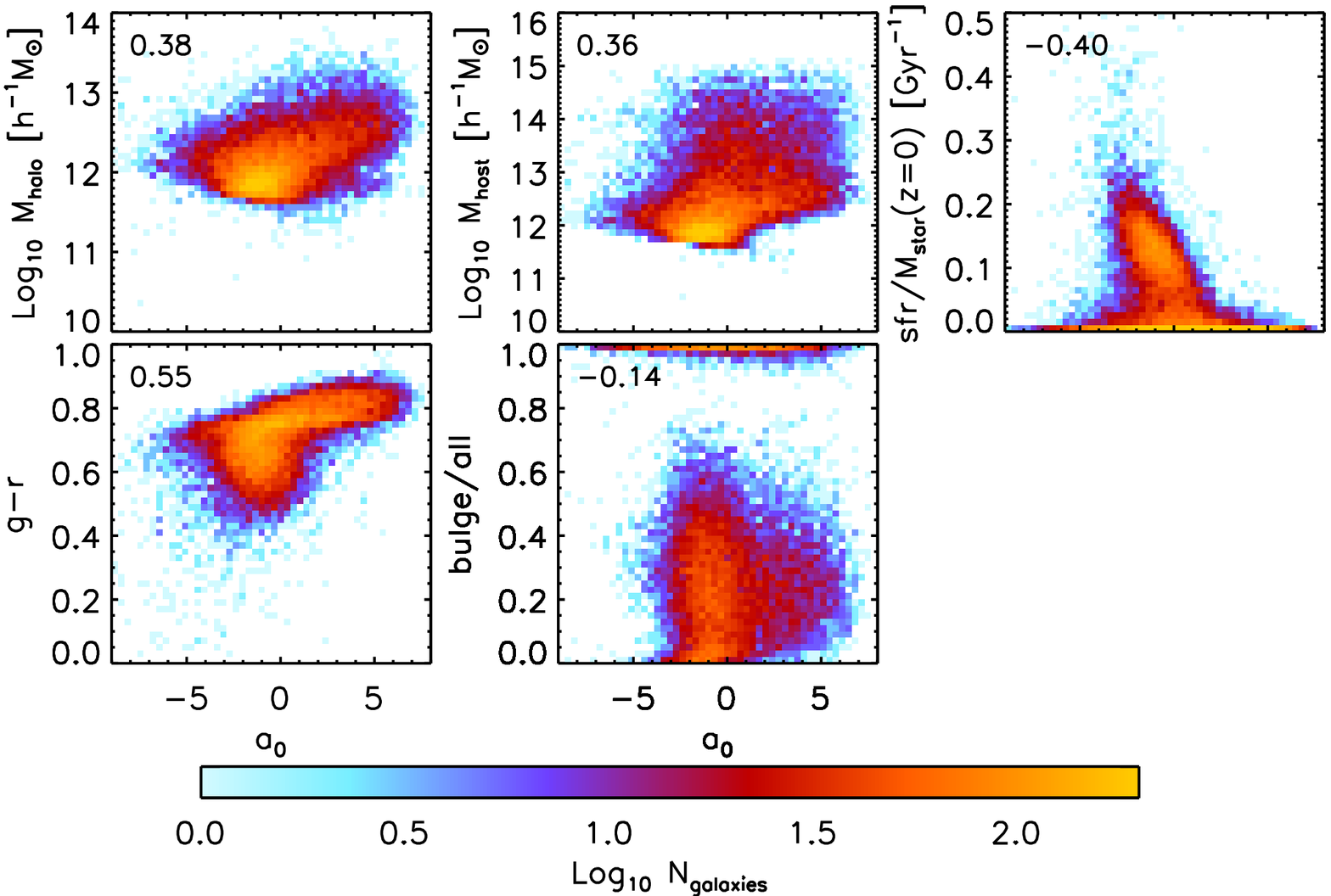}}
\resizebox{3.5in}{!}{\includegraphics{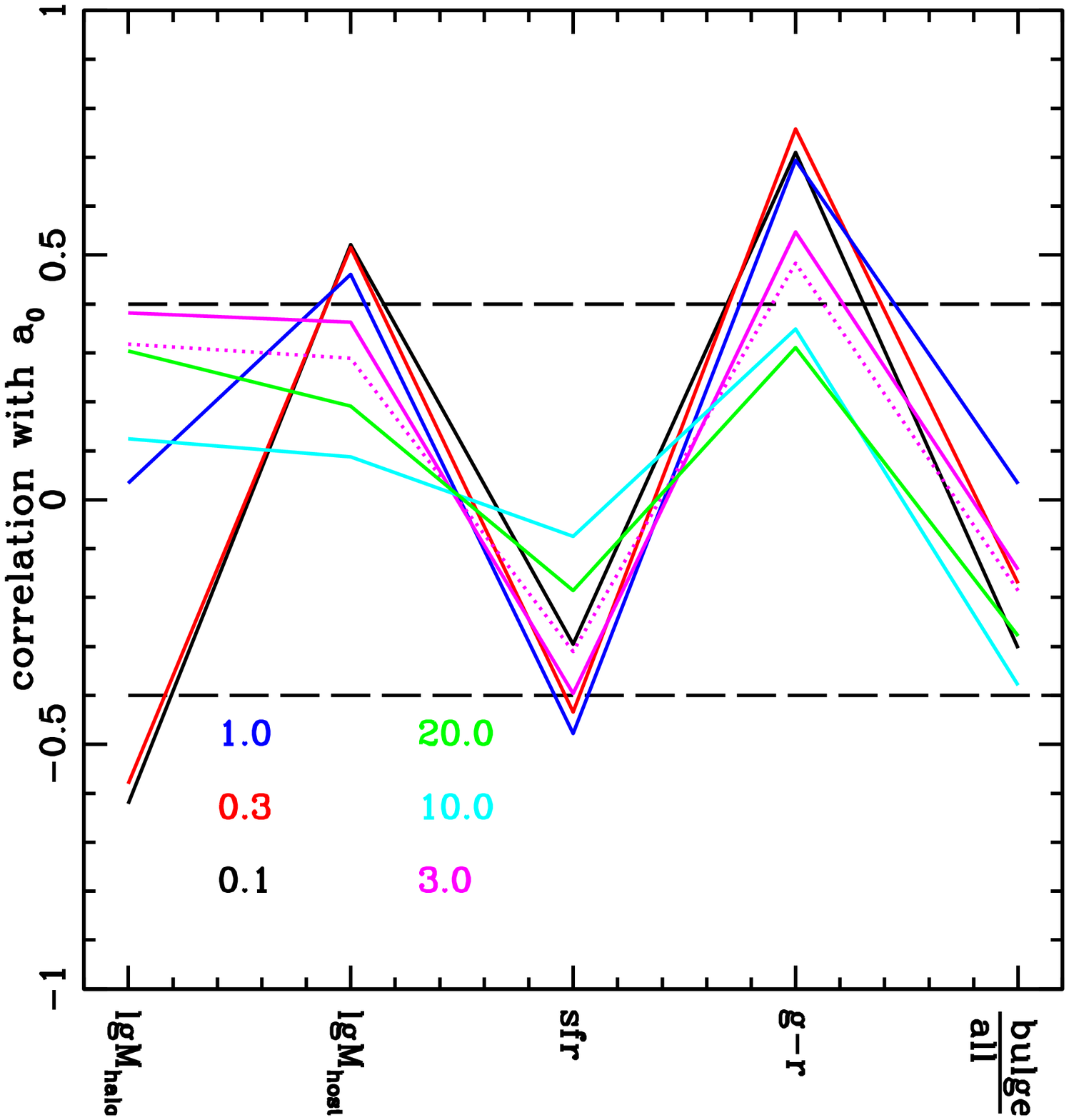}}
\end{center}
\caption{Relation of leading PCA coefficient, $a_0$, to
final time observable properties.
Top:
For fiducial sample, distribution of 
 $a_0$ relative to  $\log_{10} M_{\rm halo}$, $\log_{10} M_{\rm
  host}$,
sfr$/M_\star(z=0)^2$, $g-r$ colour, and bulge/all stellar mass.
Bottom: Correlation coefficients for all 6 Millennium samples, for the
same quantity. 
Coloured curves are for different
  masses, as indicated in the legend in terms of
$M_\star(z=0)/10^{10}M_\odot$ in the lower left corner.
The dashed line is the correlation coefficient
for
$\alpha_0$, rather than $a_0$, i.e. the signed fractional contribution to
$\Delta M^2$ from $PC_0$, rather than its value, for the fiducial
sample. We find strong correlations with halo and host mass, star
formation rate, and colour, especially for low stellar mass bins. 
}
\label{fig:obscorr}
\end{figure}

We now turn to relations between $a_0$ and several $z=0$
observables, to explore the relation between 
PCA characterizations of stellar
mass histories and possible fixed time measurements.
For final time observable quantities, we took
$\log_{10} M_{\rm halo}$, $\log_{10} M_{\rm host}$ (for a
satellite this is the mass of the main halo in which it is a satellite, rather
than its subhalo mass), star formation rate (sfr), g-r colour (SDSS, with
dust) and bulge/all stellar mass ratio.  These were all calculated or provided
directly from the Millennium \citet{Guo13} database.  
Scatter plots of these quantities versus $a_0$ are shown in the top
panel of
Fig.~\ref{fig:obscorr}, for the fiducial sample,
and correlation coefficients for all 6 Millennium samples are
shown in the bottom panel.  The correlation coefficients for $\alpha_0$ for
the
fiducial sample are also shown. 
The correlations between stellar mass history PCA properties
and final time galaxy properties may give clues as to
how to estimate stellar mass history properties from
fixed time observables.   
Many large correlations are seen, and often are as expected. For
instance, large
$a_0$ indicates a large early stellar mass gain
component and is correlated with larger g-r colour, i.e. redness and lower star
formation.  For the higher final stellar mass samples,
large $a_0$ is also correlated with higher halo mass and host halo
mass. Again,
correlations are weaker for
the two highest stellar mass samples, for the reasons given above.
For the two lowest final stellar mass samples ($\leq 0.3 \times
10^{10}M_\odot$),
$M_{\rm halo}$ and $a_0$ have a negative correlation
(satellites cause the correlation between $a_0$ and $M_{\rm host}$ to
remain positive).  
For these, then, the higher the
halo mass, the later the stellar mass gain, reversing as final stellar
mass increases. 
Another feature (not shown) is that
for all but the highest stellar mass sample, a large $\Delta M_+^2 \propto \sum_{m=3}^{N}
a_m^2$ is
strongly correlated with a large bulge/all stellar mass ratio.\footnote{Of course,
many of the final properties and halo history properties are
correlated with each other, as has been studied in several works, e.g.
\citet{WonTay11} or \citet{Che12}.}

To summarize, the sign of the contribution to $PC_0$
corresponds to a history component with early or late stellar mass growth, and
the signed fractional contribution of $PC_0$, $\alpha_0$,
is often large and negative for the observationally derived star
forming main sequence in terms of the bases found for
the Millennium
samples. 
This fractional contribution and the total contribution $a_0$ of
$PC_0$ 
in Millennium
are both strongly correlated with $z=0$ colour (g-r) and star
formation, as well as $z_{\rm starve}, z_{50}$ and other galaxy halo
and stellar mass history
times.
Increasing the bulge/all stellar mass ratio often
correlated with an increase in the total deviation from the average
history not due to the first three $PC_n$.  Again,
differences appeared between the lower four and upper two
$M_\star(z=0)$ samples, which may be in part related to the fact that
$PC_0$ is less dominant for the latter.
\clearpage
\section{Splitting up the histories}
\label{sec:splitsamp}
Although smooth trends with $a_0$ can be seen in
Figs.~\ref{fig:histcorr} and~\ref{fig:obscorr}, 
it is also clear that a separation into
different classes appears when one considers $\alpha_0$, the fraction
of a galaxy's total deviation from the average history, $\Delta M^2$,
due to $PC_0$ (Fig.~\ref{fig:pc0props}).  In this case, three
populations seem to be present, with trends between the populations implied by
the correlations in Figs.~\ref{fig:histcorr} and~\ref{fig:obscorr}.   
However, splitting up 
the sample of galaxies
according to the $PC_0$ contribution implies that galaxies with late
stellar mass gain are all related to a form of fluctuation that is the
mirror image of those of galaxies with early stellar mass gain (i.e.
both proportional to $PC_0$).

One can use a more flexible characterization to divide up galaxy
histories, by associating each galaxy with one (or more) of several
different average paths or classes, rather than with symmetric
fluctuations around a single central average history.  We considered
two ways of finding such paths, mixture models and k-means clustering.
These do not assume that different basis histories are ``orthogonal''
and thus can allow more general basis histories, if preferred by the
data.  Mixture models model the full distribution (of stellar mass
histories, in this case) with a combination of several distributions,
each here taken to be the same functional form but with different
parameters.  We took Gaussian distributions (a common choice) as our
basis distributions, each with a different average stellar mass
history and different scatters.  k-means clustering partitions objects
into sets, with each object assigned to the set with the closest
centre (in this case the closest average history\footnote{Distance
  squared is taken to be the sum over time outputs of the square of
  the difference of the stellar masses.}), but does not assume a form
for the distribution of galaxy stellar mass histories around each
centre.  In both, one has to choose the number of components and
provide initial guesses for centres.  One then solves iteratively for
better centres and corresponding galaxy assignments to each component
(and, for mixture models, for improved scatters and weights in the
full distribution).

We used mixture models based on the first 20 $PC_n$
coefficients 
$\{a_n\}$ (see Appendix for how well this approximation works in the
Millennium samples).  
Mixture models based on $M_\star(t_i)$ 
tended to become numerically unstable due to the large
powers in the 41 dimensional determinant of the Gaussian scatters.  
The initial paths were taken to be a range of  functions of the leading $PC_n$.
For the k-means clustering, each galaxy history $M_\star(t_i)$
is assigned to the $j$th average path $\mu(t_i)^j$
for which
$(M_\star(t_i)-\mu(t_i)^j)^2$ is the smallest.  Using $M_\star(t_i)$
directly was
stable for k-means clustering, so we used the full histories
(rather than the first 20 $PC_n$ coefficients as was the case for
the mixture models).
The initial $\mu(t_i)^j$, where $j$ runs over the number of 
components, were chosen randomly from the galaxy distributions; once
galaxies
are assigned to average paths, the averages are recalculated, and the
process  reiterated.

Both search methods converge to local solutions, so we
checked for robustness by finding the best parameters for 9/10 of the
sample (training sample), then testing this solution on the
remaining 1/10 (test sample), and repeating this 9 times.  We found 
that increasing the number of average paths tended to increase the
likelihood of the test sample, even up to $\sim 13$ basis
histories (the only exception was a hint of an upturn in the likelihood
for
the  $10^{11} M_\star(z=0)$ sample,
in
one approximation, with 13 components).  However, with increasing numbers of components, the
interpretations of the resulting sub-populations became more
complicated.  Although it is possible that the separate classes might be distinguishing
between further galaxy properties,
given the relative success of the PCA description (galaxy histories
tend to be centred at the average history, with a large fraction of
the variance in the first fluctuation),
we focussed primarily on the 3 component models for further study.

For the lowest $M_\star(z=0)=0.1\times 10^{10} M_\odot$ final stellar
mass sample, the preferred k-means and Gaussian mixture model average
paths coincide for both 3 and 4 components.  As final stellar mass
increases, the mixture models often find average paths which
overlap significantly with each other but have different scatter, suggesting
that the Gaussian shape is not optimal (as the procedure is then
approximating a distribution 
with a narrow peak plus a large tail).  Associating a galaxy with
a large or small width Gaussian seemed
a less useful classification than
identifying the closest average history, as found using k-means.
We thus report the k-means results in more depth below.

\begin{figure}
\begin{center} 
\resizebox{3.5in}{!}{\includegraphics{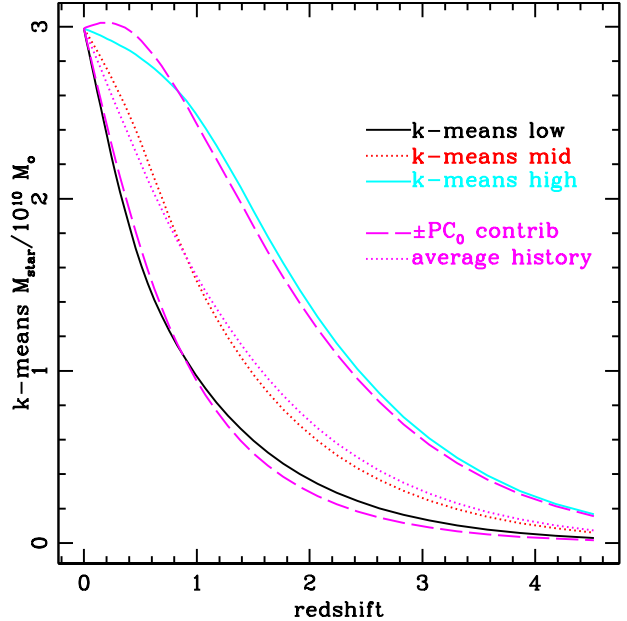}}
\end{center}
\caption{k-means average paths for $M_\star(z=0)\sim 3.0 \times
  10^{10} M_\odot$ sample (solid black, dotted red, solid cyan).  
Also shown are the $\pm PC_0$ (top and
  bottom dashed magenta curves) contributions to the high and low curves and
  the average history (central dotted magenta curve).  These trends occurred
  for all samples, that is, the three k-means paths are very close to $\pm
  PC_0$ and the average path respectively.
}
\label{fig:kpaths}
\end{figure}  

\begin{figure*}
\begin{center} 
\resizebox{6in}{!}{\includegraphics{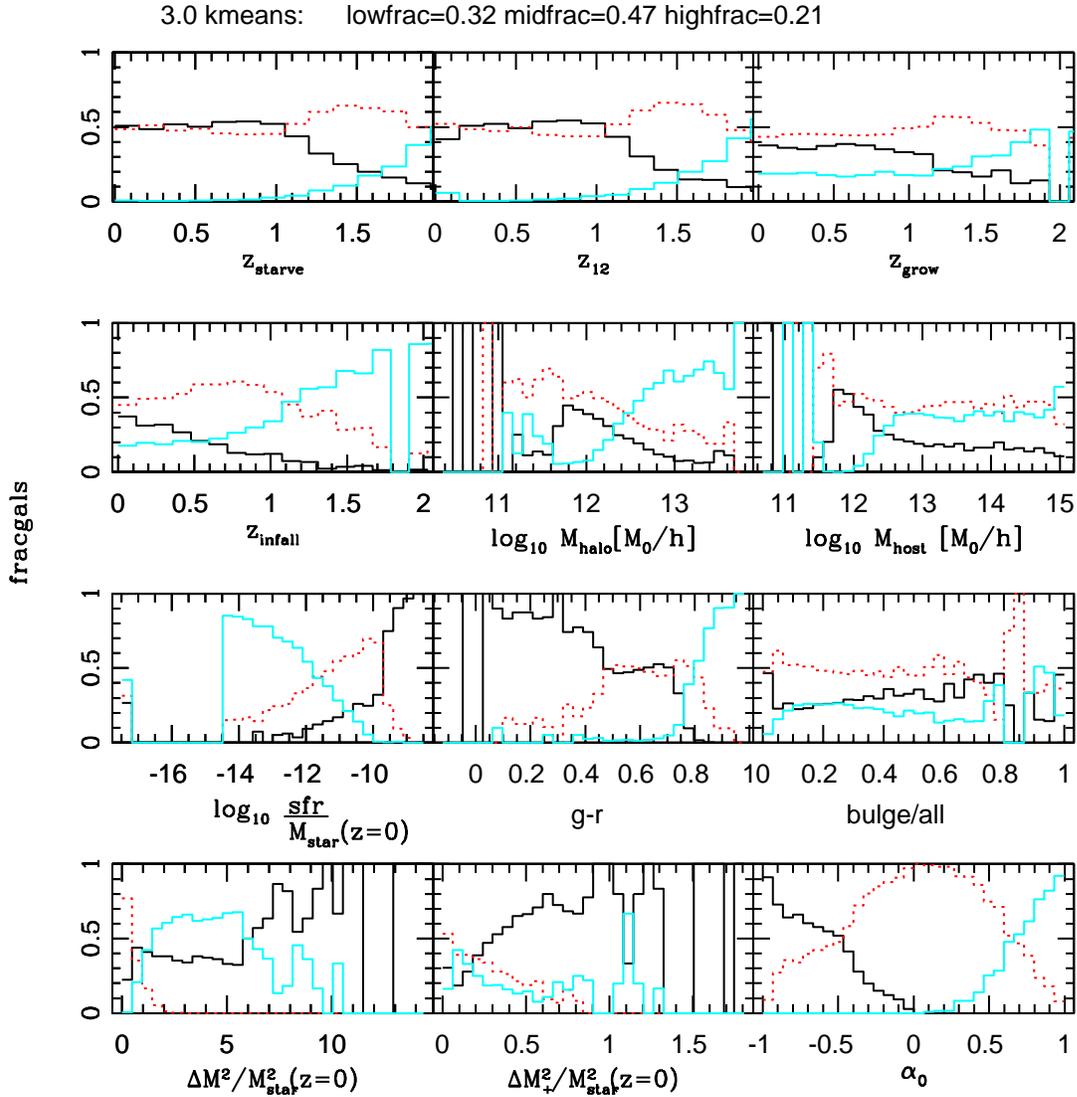}}
\end{center}
\caption{Properties of
  galaxies in each of the three k-means distributions in
  Fig.~\ref{fig:kpaths}; 
line
  types and colours of each subsample match that of their corresponding
  k-means path.  Each histogram is the fraction
  of galaxies in each k-means component, bin by bin (so the sum in any
  bin of all three components is equal to 1).
Properties defined in \S \ref{sec:corrhist} and \ref{sec:corrfinal}, similar to
Figs.~\ref{fig:histcorr} and \ref{fig:obscorr}
(star formation rate is shown logarithmically and, when zero, shown as $<-16$).  This indicates how well a given individual final time
observation
or halo history time
can be used to infer to
which k-means set a galaxy is likely to belong.
}
\label{fig:pathsplit}
\end{figure*}  

For the fiducial sample, the 3 component k-means average
histories are shown in Fig.~\ref{fig:kpaths}.
The black solid, red dotted, and cyan solid lines correspond to the
k-means average
paths, the $\mu(t_i)^j,j=1,2,3$.  Individual k-means average paths have a large overlap with
the average history (dotted magenta) or
$\pm PC_0$ (dashed magenta).
The $\alpha_0$ of the upper and lower k-means histoires
has absolute value $\geq$96\%   
for all samples. 
In addition, on average 
over 90\% of the galaxies in the lowest (highest) k-means group have
$\alpha_0 < -0.5$ ($\alpha_0 >0.5$).  Smaller corresponding fractions occur for
$a_0 < -rms(a_0)$ and $a_0 > rms(a_0)$ but the trends are the same.

Thus the $\pm PC_0$ fluctuations actually do seem to be good basis
vectors for classifying the full set of histories, even when an
orthogonal basis is not imposed.  
However, for the 4 lower final
stellar mass samples, the k-means history 
associated with $+PC_0$ has a coefficient $a_0$ 
much larger than the coefficient $a_0$ for the k-means history 
associated with $-PC_0$, with the trend reversing for 
the 2 highest final stellar mass samples.  That is, the early
and late stellar mass gain k-means average histories, although both related to
$\pm PC_0$, are not direct reflections of each other.  The fraction of
galaxies in the lowest and highest k-means histories, for all samples,
is listed in Table \ref{tab:samprops} under ``lowk''/``highk''; changes are
seen with final stellar mass, but they were not simple to interpret.
For the OWLS samples again galaxies which lost more than 50\% of their
stellar mass on one time step were not included in finding the k-means
average paths.  These classifications, in terms of basis paths and the
fraction of galaxies associated with each path, may be further ways
to distinguish and compare models.  Note that ``lowpc''/``highpc'' counts
galaxies
which have more than half of their $\Delta M^2$ due to $PC_0$, i.e. it
is a property of how much of their deviation is due to $PC_0$, while
``lowk''/``highk''
distinguishes galaxies whose total deviation is further than a certain
amount from the average path, irrespective of which fluctuations contribute.

We next turn to final time, i.e.~$z=0$, and halo history properties 
in the context of k-means classification.
In Fig.~\ref{fig:pathsplit}, we take the three k-means groups
of galaxies from Fig.~\ref{fig:kpaths} and look at their relative
contributions to the population of galaxies, when considering specific
final time or halo history properties.
(The line types and colours of the subsamples in
Fig.~\ref{fig:pathsplit} 
correspond to the paths in
Fig.~\ref{fig:kpaths} with which they are associated.)  For each quantity, in each bin, the
fractional contribution of galaxies associated with each k-means path
is shown, i.e. contributions from all three
k-means subsets add up to one, in each bin.  In some cases (for
example very high star formation rate), the likelihood of being in a
particular k-means set is very high, but for other measurements (for
example bulge/all ratio),
histories can belong to several of the different k-means sets and yet
yield very similar $z=0$ observables.  The other stellar
mass samples had similar behaviour, although for lower final stellar mass there
were sometimes trends visible in the bulge/all contributions, and the
trends with $\log M_{\rm halo}$ varied for the next highest final
stellar mass sample (the highest final stellar mass sample is very
small, leading to a very noisy histogram).  Using a split based upon
$\alpha_0$ or $a_0$ gives similar results, as expected from the
overlap of the populations mentioned above.

To summarize, splitting up galaxy histories into 3 sets tended to
give a set associated with
the average history of the full sample, and sets with 
early and late stellar mass gain histories, whose average paths significantly
overlap with the leading PCA fluctuation $PC_0$.  The separation into the
three sets was similar to that implied by using contributions to the
galaxy's history from $PC_0$.  We identified how
well specific values of some particular $z=0$
observables or halo history characteristic times implied a galaxy
belonged to a particular k-means average path.  For some particular measured
values of certain galaxy
properties, the corresponding galaxy's stellar mass history
was highly likely to lie within one particular k-means subsample.

\section{Summary and Discussion}
\label{sec:discussion}

We applied a standard classification tool, principal component
analysis (PCA), to several collections of simulated stellar mass histories
from the Millennium simulation and from the OWLS project.
This approach characterizes the full population of galaxy histories
ending at a given stellar mass, as well as giving a new description of
each galaxy history in the sample.  As such, it provides quantitative
and qualitative methods to describe ``how galaxies form'' (in a given
model), 
starting with the actual simulated galaxy histories.

We considered simulated stellar mass histories 
for several different final stellar mass
($M_\star(z=0)$) galaxy samples.  For each, the distribution of
histories
tended to peak around a central value and one main
perturbation (change in history from average) dominated the variance
(scatter) of the ensemble of galaxy histories from the average.
Combining the
fluctuations due to the first three
leading perturbations accounted for
$\sim$90\% of the variance around the average stellar mass history.
Similar trends were seen for the halo mass histories, and the leading
PCA halo history and stellar mass history contributions
were correlated for most of the Millennium final stellar mass
samples.
Star
formation rate histories did not have the leading principal components
dominate, while fractions of the variance due to 
specific star formation rate history principal components fall in between those for star
formation rate and stellar mass for the first $\sim$20 components, then
drop below both.

 In terms of individual galaxies, the leading perturbation
from the average history 
corresponds to a particular form of either late (leading perturbation with a minus sign) or
early (with a plus sign) stellar mass gain, relative to the average
stellar mass history.  
The shape of the leading fluctuation 
is similar amongst the Millennium
lower final stellar mass samples
($\leq 3 \times 10^{10}M_\odot$) and the OWLS AGN+SN sample,
and between the two Millennium highest final
stellar mass samples and the OWLS SN only simulation.
Thus this component can vary with changes in physics, 
and with final stellar masses within the same simulation.
This characteristic property of a set of stellar mass histories
can perhaps be used to intercompare them.

This leading principal component also comprises a large fraction of
many individual galaxy histories. We quantified how well using a given
number of PCA components approximates the Millennium sample galaxy
histories in the Appendix.

Given the importance of the leading perturbation to the average
history, we searched for and found correlations between its
coefficient and several characteristic stellar mass history times,
halo history times and final time observables in the Millennium
samples. We found in particular strong correlations with g-r colour
and instantaneous star formation rate for the lower final stellar mass
samples.  For all but the highest final stellar mass sample, galaxies
with large bulge/stellar mass ratios tended to have larger
fluctuations from the average path which were not due to the leading
few PCA components.  All Millennium samples had large correlations of
their amount of leading principal component with halo history times
such as $z_{\rm starve}$.  The leading principal component also had,
not surprisingly, large correlations with times where the stellar mass
history reached some fixed fraction of its final value.

Going beyond a single average history plus its perturbations, 
we separated the galaxy histories into 3 classes using the
contribution
from the leading PCA component, mixture models
and k-means clustering.  The resulting divisions of
galaxies
often were similar in terms of the resulting basis paths
and in terms of the distribution of galaxies with a
certain
observable property.  For instance, galaxies which are very red (large
g-r)
almost all have at least 50\% of their $\Delta M^2$, their separation squared
from the average path,  due
to $+PC_0$ (i.e. $\alpha_0>0.5$).
This helps in gaining intuition for the
classification found by these three methods and gives an
idea of how well one can decode a galaxy's stellar mass history
given these fixed time observations.  The fractions of
galaxies in these different classes changes between
different final stellar masses and different models, and 
may also provide a point of
comparison between them.

All of these tools and quantities provide several ways to look at a
full 
ensemble of
galaxy stellar mass histories, such as are being produced by many
different methods and groups.   These ensembles of full histories
answer ``how do galaxies form'' in any given model, by
incorporating the consequences of all the modeling assumptions used
to create them.
Going forward, the PCA classification provides
a simple characterization of the full set of galaxy histories, both in
finding how much different perturbations contribute to the histories,
and in the form of the perturbations themselves.
Histories generated with different methods or having different final
stellar
mass sometimes, but not always, exhibited different
properties.  
The principal components 
can also be used to approximate galaxy histories using a small number of basis
components.  All of these features 
might be useful when exploring simplifying
assumptions and models for galaxy formation.  One could ask
how these basis components, and the galaxy contributions from them,
change as the models change. 
This is particularly straightforward
in semi-analytic models (including highly simplified versions such as
\citealt{MutCroPoo13}, \citealt{Lu14}, and \citealt{Bir14}), where the dark matter remains unchanged.
Differences between families of histories, using this
classification, could be easily generated and compared. 

More specifically, quantities for comparison include the total
variance, the (fractional) variance due to $PC_0$ and in the leading
components, the shape of $PC_0$ or other $PC_n$, the number of galaxies with large or
small, raw or fractional contributions from $PC_0$, the average paths
found using k-means clustering or mixture models, the distributions of
galaxies into the k-means sets, and their counterparts for say central
or satellite galaxies, and other subsamples. 
Analogues for other histories such as halo
mass (although subhalos cause complications) and specific star
formation rate may also be of use.
 Once a family of
stellar mass histories is in hand, calculation of the principal
components and the k-means sets is straightforward and quick
(mixture models are more complicated in part because they encode
more information, in particular the shape of the distribution around
the average histories).  

One further step is to tie these descriptions of histories more closely to
observations.  
For simulations which report observable properties, these stellar mass history
classifications provide points of contact between them and the full
set of galaxy stellar mass histories.
(The observable properties we
considered were not comprehensive, using more colours, for instance is
a direct generalization.)   We explored these briefly, including the
effects of changing final stellar mass.  Changing simulation
methods or assumptions can be used to identify
which properties of the ensemble of histories, such as those
identified here, are most connected to
which observational properties.

\section*{Acknowledgements}

JDC thanks S. Axelrod, D. Cohn, N. Dalal, D. Freed, S. Genel, S. Ho,
S. Leitner, G. Lemson and B. Sherwin
for suggestions and discussions, and especially thanks M. White for
numerous suggestions and discussions.  We thank A. Hearin and M. White
for comments
on an earlier version of the draft.
JDC is supported in part by DOE.
We thank the OWLS team for the use of the simulations.
The OWLS simulations were run on Stella, the LOFAR
BlueGene/L system in Groningen, on the Cosmology Machine at the
Institute for Computational Cosmology in Durham as part of the Virgo
Consortium research programme, and on Darwin in Cambridge. 
This work was supported in part by National Science Foundation Grant No. PHYS-1066293 and the hospitality of the Aspen Center for Physics.
 JDC also thanks the Royal
Observatory of
Edinburgh and the Max-Planck-Institut f\"ur Astronomie for hospitality
while completing this work.

\section*{Appendix: Approximating galaxy histories by a subset of $PC_n$}
\label{sec:approx}
As the leading few $PC_n$ capture a large part of the variance of each
sample relative to the average history, one can also ask how well
using only a small number of $PC_n$ can approximate individual galaxy
histories.  We focus on the 6 Millennium samples (in part for
simplicity
and in part because the different number of output times makes
comparisons with the OWLS models less direct).

\begin{figure}
\begin{center} 
\resizebox{3.5in}{!}{\includegraphics{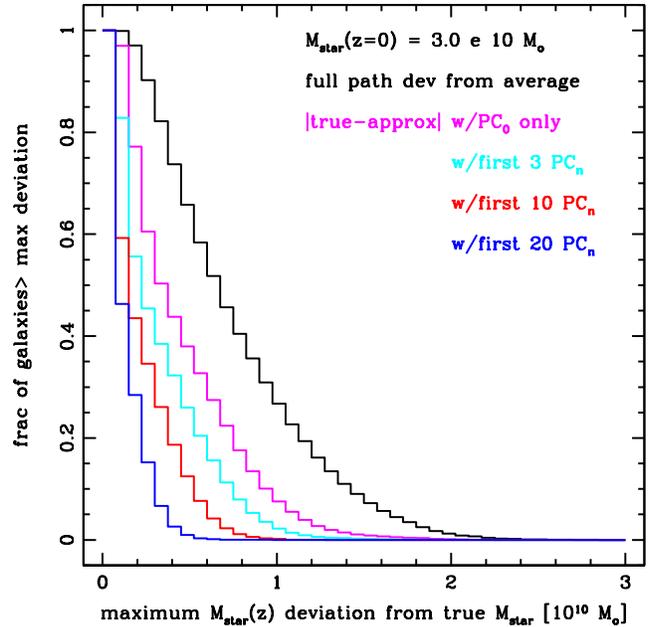}}
\end{center}
\caption{
For the fiducial sample, the distribution of the maximum
deviation of all galaxies from their true history (at any one given time $t_i$)
from several different approximations:  using the average of all
galaxies (black histogram), only $PC_0$ (magenta histogram), and the first 3, 10, and 
 20 $PC_n$ (cyan, red, and blue histograms, respectively).  Using a few $PC_n$ quickly brings many galaxy histories
 close to their true values.
}
\label{fig:pckeepgal}
\end{figure}
 In Fig.~\ref{fig:pckeepgal}  we 
show the distribution of the largest deviation between true and approximate galaxy
histories, at any one time $t_i$, for all of the galaxies in the fiducial sample.
Five approximations are shown:
$M_\star(t_i) = $ the average history plus 0, first 1, first 3, first 10 or
first 20 $PC_n$ contributions.  Most of the histories become quite
close to their true histories by the time 20 (less than half) of the 
$PC_n$ are used.
\begin{figure}
\begin{center} 
\resizebox{3.5in}{!}{\includegraphics{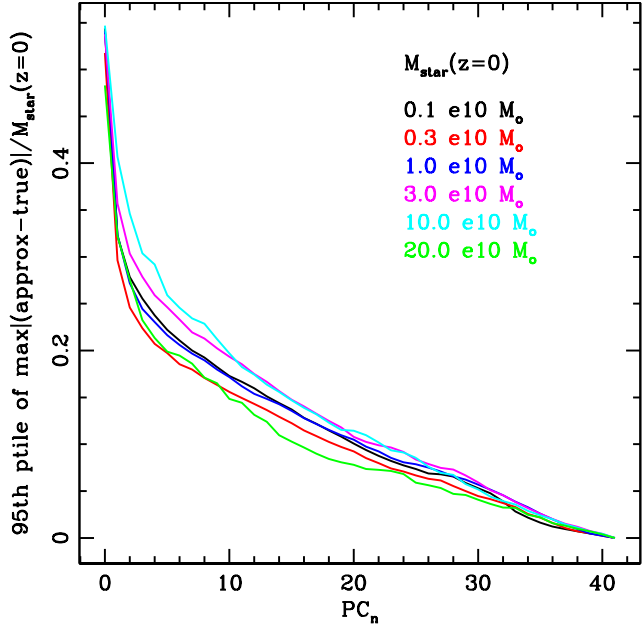}}
\end{center}
\caption{Approximating galaxy histories by a subset of $PC_n$:
Statistics for all 6 Millennium samples: the 95th percentile
of the 
maximum deviations at any given $t_i$ in a galaxy history, when keeping
only the first $n$ $PC_n$.  Fig. ~\ref{fig:pckeepgal} gives an example
of the distribution below the 95th percentile for going up to
$PC_0$,$PC_3,PC_{10}$ and $PC_{20}$ for the fiducial sample.
The $y-axis$ value is rescaled by each sample's
final stellar mass to show general trends and allow more direct 
comparison.  
}
\label{fig:distmax}
\end{figure}

\begin{figure}
\begin{center} 
\resizebox{3.5in}{!}{\includegraphics{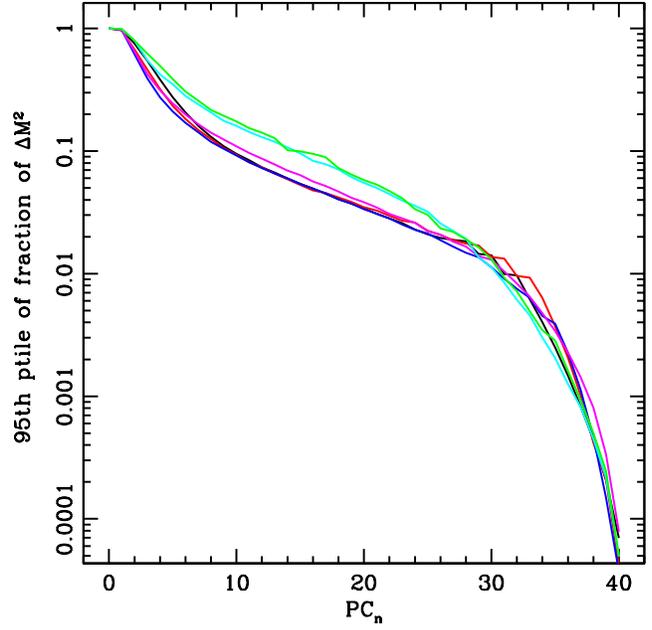}}
\end{center}
\caption{
95th percentile of fraction of the variance due to all $PC_m$
components,
above $m \geq n$,
i.e. keeping only up to and not including the $n$th $PC_n$ will omit this fraction of the
variance or less for 95\% of the galaxies, again for all Millennium
samples.  Colours as in Fig.~\ref{fig:distmax} above.
}
\label{fig:distfrac}
\end{figure}
To show trends with sample and with keeping increasing numbers of
$PC_n$, the 95 percentile of maximum deviations ($|true-approximate|$
history at any given time) are shown in Fig.~\ref{fig:distmax}, for
all 6 Millennium samples.  The deviations are rescaled by the final
stellar mass, to allow comparison between different samples and to
highlight the similarities.  The 95th percentile maximum deviation
along a history can be large.  However, the median largest deviation
of each sample, for galaxy histories approximated at least up to
$PC_{10}$, was at or below 16\% of the 95\% largest deviation.
(Again, the shape of the distribution of large deviations for all
galaxies is shown for the fiducial case in Fig.~\ref{fig:pckeepgal}
above.)  
For $\Delta M^2$, approximately 10 $PC_n$ will get most galaxies in
the four lower stellar mass samples within 10\% of their true
histories, a few more $PC_n$ are needed for the two highest final
stellar mass samples.  This can be seen in Fig.~\ref{fig:distfrac}
bottom, showing the 95th percentile of the maximum $\Delta M^2$
difference between the true and approximate histories, when dropping
a given $PC_n$ and above.  This quantity pertains to the
individual galaxy histories, although it is related to
the total variance in the sample due to the respective principal
components.

\end{document}